\documentclass[12pt]{article}
\usepackage[utf8]{inputenc}
\usepackage{geometry}
\usepackage{graphicx}
\usepackage{array}
\usepackage{subfig}
\usepackage{slashed}
\usepackage{amsmath}
\usepackage{amsfonts}
\usepackage{amssymb}
\usepackage[cm]{fullpage}
\usepackage{cite}
\usepackage{epstopdf}
\usepackage{comment}
\usepackage{hyperref}
\usepackage{cancel}
\usepackage{mathrsfs}

\usepackage{cleveref}
\crefname{section}{§}{§§}
\Crefname{section}{§}{§§}

\usepackage{booktabs} 

\numberwithin{equation}{section}

\def\p{\partial}

\def\0{{(0)}}
\def\1{{(1)}}
\def\2{{(2)}}

\def\<{\langle }
\def\>{\rangle }

\newcommand{\bea}{\begin{eqnarray}}

\newcommand{\eea}{\end{eqnarray}}

\newcommand{\be}{\begin{equation}}
\newcommand{\ee}{\end{equation}}
\newcommand{\ba}{\begin{align}}
\newcommand{\ea}{\end{align}}

   \makeatletter
  \let\over=\@@over \let\overwithdelims=\@@overwithdelims
  \let\atop=\@@atop \let\atopwithdelims=\@@atopwithdelims
  \let\above=\@@above \let\abovewithdelims=\@@abovewithdelims
\renewcommand\section{\@startsection {section}{1}{\z@}%
                                   {-3.5ex \@plus -1ex \@minus -.2ex}
                                   {2.3ex \@plus.2ex}%
                                   {\normalfont\large\bfseries}}

\renewcommand\subsection{\@startsection{subsection}{2}{\z@}%
                                     {-3.25ex\@plus -1ex \@minus -.2ex}%
                                     {1.5ex \@plus .2ex}%
                                     {\normalfont\bfseries}}

\newcommand{\beq}{\begin{equation}}
\newcommand{\eeq}{\end{equation}}
\newcommand{\beqa}{\begin{eqnarray}}
\newcommand{\eeqa}{\end{eqnarray}}
\newcommand{\beqar}{\begin{eqnarray*}}

\def\[{\[}
\def\]{\]}

\newcommand{\bd}[1]{\begin{fmffile}{#1}\begin{fmfgraph*}}
\newcommand{\ed}{\end{fmfgraph*}\end{fmffile}}

\begin{document}

\begin{titlepage}

\begin{flushright}
{\small LMU--ASC 70/17\,\, }\\
{\small  MPP--2017--254\,\,  }\\
{\small CERN-TH-2017-263}
\vspace{1cm}
\end{flushright}

\unitlength = 1mm~\\
\begin{center}

{\LARGE{\textsc{2d orbifolds with exotic supersymmetry }}}

\vspace{1cm}
{\large Ioannis Florakis$^a$,\ \  I\~naki Garc\'ia-Etxebarria$^b$,}
\\
{~}\\
{\large  Dieter L\"ust$^{b,c}$,\ \ Diego Regalado$^d$  }

\vspace{1cm}

\centerline{\it $^{\textrm{a}}$ Laboratoire de Physique Th\'eorique et Hautes Energies}
\centerline{\it CNRS UMR 7589 - UPMC Paris VI, 4 Place Jussieu, 75005 Paris, France}
\medskip
\centerline{\it $^{\textrm{b}}$ Max--Planck--Institut f\"ur Physik,
Werner--Heisenberg--Institut,}
\centerline{\it 80805 M\"unchen, Germany}
\medskip
\centerline{\it $^{\textrm{c}}$ Arnold--Sommerfeld--Center for Theoretical Physics,}
\centerline{\it Ludwig--Maximilians--Universit\"at, 80333 M\"unchen, Germany}
\medskip
\centerline{\it $^{\textrm{d}}$ Theoretical Physics Department, CERN, Geneva, Switzerland}

\vspace{0.8cm}

\begin{abstract}

  We analyse various two dimensional theories arising from
  compactification of type II and heterotic string theory on
  asymmetric orbifolds. We find extra supersymmetry generators arising
  from twisted sectors, giving rise to exotic supersymmetry algebras.
  Among others we discover new cases with a large number of supercharges, such as $\mathcal{N}=(20,8)$, $\mathcal{N}=(24,8)$,
  $\mathcal{N}=(32,0)$, $\mathcal{N}=(24,24)$ and
  $\mathcal{N}=(48,0)$.

\end{abstract}

\setcounter{footnote}{0}

\vspace{1.0cm}

\end{center}

\end{titlepage}

\pagestyle{empty}
\pagestyle{plain}

\def\vx{{\vec x}}
\def\p{\partial}
\def\po{$\cal P_O$}

\pagenumbering{arabic}

\tableofcontents

\newpage


\section{Introduction}

Supersymmetry is undoubtedly one of the main ingredients of string
theory. Although, historically, the development of string theory and
supergravity originated from somewhat different paths, the strong
connection between the two has become essential for our understanding
and investigating their rich structure.  Over the last four decades,
supergravity theories \cite{Freedman:1976xh,Deser:1976eh} in $D\geq4$
dimensions have been studied in great detail (for overviews see, for
instance, \cite{sugra,Freedman:2012zz}).  Moreover, this discussion
has been extended in \cite{deWit:1992psp}, where a full classification
of all possible number of supercharges has been obtained for the case
of $D=3$.  The main purpose of the present work is to fill a gap in
the landscape of $D=2$ extended supergravity theories by investigating
superstring vacua with an ``exotic" number of
supersymmetries. Extending previous results
\cite{Sen:1996na,Dasgupta:1996yh,Font:2004et} we will present novel
two-dimensional string constructions obtained as asymmetric
$T^8/{\mathbb Z}_n$ or $T^8/{\mathbb Z}_n\times \mathbb Z_m$ orbifold
compactifications \cite{Dixon:1985jw,Narain:1986qm} with chiral
$({\cal P},{\cal Q})$ supersymmetry, namely, with ${\cal P}$ positive
and ${\cal Q}$ negative helicity supercharges.

In particular, we will construct new asymmetric ${\mathbb Z}_4$
orbifold compactifications\footnote{Similar left-right asymmetric
  string constructions were considered in the context of non-geometric flux compactifications
  \cite{CFL,CFKL}, from the point of view of exactly solvable worldsheet CFTs, realized as freely-acting asymmetric orbifolds.
Closely related theories were also obtained in the framework of  double  field theory in \cite{Hassler:2014sba}, while 
 generalizations realised as Gepner models in four, six and eight dimensions were presented in
  \cite{Blumenhagen:2016axv,Blumenhagen:2016rof}.} of type IIA, IIB or
heterotic string theory in which a very interesting phenomenon
occurs. As is well known from the familiar case of four dimensions,
requiring the absence of massless states of spin higher than two,
constrains the maximal number of supersymmetries of the theory to
$\mathcal N\leq 8$. Despite the absence of such a no-go theorem in the
more exotic case of two spacetime dimensions  no
examples of consistent 2d string or supergravity theories with more
than 32 supercharges have been presented in the literature. As we will
show explicitly in this work, twisted sectors of asymmetric orbifolds
in 2d may actually give rise to additional supercharges and, hence,
lead to consistent string theories with more than 32
supercharges. These are, to our knowledge, the first examples of such
theories in the literature.

\medskip

We have organized this work as follows. 
In \S\ref{sec:review} we will recall the structure of the chiral
$({\cal P},{\cal Q})$ supersymmetry algebra in two space-time dimensions
and emphasize that for chiral theories the massless spectrum of the
positive and negative helicity states is different. This fact is
especially prominent for chiral models with $({\cal P},0)$
supersymmetry, where the negative helicity sector is
non-supersymmetric with a different number of space-time bosons and
fermions.  In \S\ref{sec:symmetric-Z4} we discuss the type II and
Heterotic strings on a symmetric $\mathbb{Z}_4$ orbifold of
$T^8$. This section is mostly a systematic elaboration on known
results, but is nevertheless useful as an introduction to the formalism. In
\S\ref{sec:asymmetric-Z4} we move on to the analysis of  asymmetric
$\mathbb{Z}_4$ orbifolds, where we encounter our first examples of
exotic superalgebras. In \S\ref{sec:more-than-32} we analyze various
asymmetric $\mathbb{Z}_4\times\mathbb{Z}_4$ orbifolds with the
peculiarity that the resulting two dimensional spacetime theory has more than 32
gravitinos. The appendices contain various technical results we use in
the main text.

\section{Chiral  $({\cal P},{\cal Q})$ theories in two dimensions}
\label{sec:review}

Here let us recall some basic facts of supersymmetric field theories
in two dimensions and their supersymmetry algebras
\cite{Witten:1993yc,Hull:1985jv}.  First we introduce lightcone
coordinates $Y^\pm={1\over \sqrt 2}(Y^0\pm Y^1)$ in the (1+1)
dimensional uncompactified Minkowski space-time. Then a massless
particle $\Phi_+(Y^+)$ or $\Phi_-(Y^+)$ is either left- or right
moving in target space with either $P_+=0$ or $P_-=0$.  In particular,
massless spin-1/2 fermions $\Psi_+$ or $\Psi_-$ can have helicity $+1/2$ or
$-1/2$ in two space-dimensions. For the supercharges, this possibility
is made manifest by representing the supersymmetry algebra by
${\cal P}$ positive chirality supercharges, denoted by $Q_+^I$
($I=1,\dots ,{\cal P}$) and ${\cal Q}$ negative chirality supercharges
$Q_-^{I'}$ ($I'=1,\dots ,{\cal Q}$). The extended
$({\cal P},{\cal Q})$ supersymmetry algebra then takes the following form:
 \begin{equation}
 \lbrace Q_+^I, Q_+^J\rbrace=2\delta^{IJ}P_+\, ,\quad  \lbrace Q_-^{I'}, Q_-^{J'}\rbrace=2\delta^{I'J'}P_-\, ,\quad \lbrace Q_+^I, Q_-^{J'}\rbrace=0\, .
 \end{equation}

In string theory one can represent the supersymmetry algebra in terms of the vertex operators of the supercharges (see e.g. \cite{Ferrara:1989ud}).
Specifically in ten space-time dimensions, 
space-time supersymmetry transformations in type II (heterotic) string theories are generated by 32 (respectively 16)
supercharges, coming from  left- and right-moving sectors of
the world-sheet  superconformal field theory. 

Upon compactifying, the number of preserved supercharges is generically reduced. However, in later sections, we shall be considering asymmetric 2d orbifolds with additional supercharges arising from the twisted sectors, such that the total number of supercharges can be even higher than in the ten-dimensional theory.
Consider first the left-moving supercharges, which are identical in type IIA,B and heterotic string theories.
All $({\cal P}_L,{\cal Q}_L)$ space-time supercharges in two dimensions
are realized on the world-sheet by covariant weight-1 vertex operators that necessarily contain the spin fields $S_\pm(z)=e^{\pm {iH_0(z)\over 2}}$. Here, $\psi_\pm=(\psi_0\pm i\psi^1)/\sqrt{2}=e^{\pm {iH_{0}(z)}}$ is the  
complex world-sheet fermion in the light-cone directions, that may be bosonized in terms of the current
 $J_0(z) = i\partial H_0$.
In addition, we need the 
internal spin fields $S^I(z)$ ($I=1,\dots ,{\cal P}_L$) and $S^{I'}(z)$ ($I'=1,\dots ,{\cal Q}_L$) with conformal dimension $h=1/2$ of  the internal eight-dimensional space. In terms of these, the vertex operators of the supercharges in the (non-canonical) $+1/2$ ghost picture  take the  form:
\begin{eqnarray}\label{supercharges2drightge}
Q_+^{  I} &=&\oint {dz\over 2\pi i}S^{I}(z)e^{{\phi(z)+iH_{0}(z)\over 2}} 
	 \,,\nonumber\\
Q_-^{  I'} &=&\oint {dz\over 2\pi i}S^{I'}(z)e^{{\phi(z)-iH_{0}(z)\over 2}} 
	 \, .
\end{eqnarray}
Here, $\phi(z)$ is the bosonised superghost field.  

Upon dimensional reduction from ten to two dimensions on  $T^8$, the 16 right moving supercharges transform in the 16-dimensional spinor
representation\footnote{We choose a convention where the SO(10) spinor weights $(\pm\frac{1}{2},\pm\frac{1}{2},\pm\frac{1}{2},\pm\frac{1}{2},\pm\frac{1}{2})$ with  even (resp. odd) number of minus signs correspond to the $\textbf{16}_c$ (resp. $\textbf{16}_s$) representation.} ${\bf{16}}_c$ of $SO(10)$, which decomposes under $SO(2)\times SO(8)$ as 
\begin{equation}\label{decomp}
{\bf{16}}_c^L \rightarrow (+\tfrac{1}{2},{\bf {8}}_s)^L+  (-\tfrac{1}{2},{\bf{8}}_c)^L\, .
\end{equation}
The IIA, IIB and  heterotic vertex operators of the  corresponding 16  supercharges $Q_+^I$ and  $Q_-^{I'}$  with $({\cal Q}_L,{\cal P}_L)=(8,8)$
are explicitly:
\begin{eqnarray}\label{supercharges2dright}
Q_+^{  I} &=&
	\oint {dz\over 2\pi i}e^{i w^{ I}_s\cdot  H(z)}e^{{\phi(z)+iH_{0}(z)\over 2}} \,,\nonumber\\
Q_-^{  I'} &=&
	\oint {dz\over 2\pi i}e^{i  w^{ I'}_c\cdot H(z)}e^{{\phi(z)-iH_{0}(z)\over 2}} \, .
\end{eqnarray}
where the four chiral bosons $H_K(z)$ originate from the bosonization of the internal complexified   world-sheet fermions $\psi_K$.
The eight  different 4-vectors $ w^{ I}_s=(\pm {1\over 2},\pm {1\over 2},\pm {1\over 2},\pm {1\over 2})$ (even number of minus signs) are the spinorial weight vectors of the $\textbf{8}_s^L$
spinor representation of the internal automorphism group $SO(8)$. 
Similarly the eight  4-vectors $ w^{ I'}_c=(\pm {1\over 2},\pm {1\over 2},\pm {1\over 2},\pm {1\over 2})$ (odd number of minus signs) are the spinorial weight vectors of the $\textbf{8}_c^L$
spinor representation.

For type IIA and IIB string theories, there is also a certain number $({\cal Q}_R,{\cal P}_R)$ of supercharges originating from the right-moving world sheet sector.
In type IIB, the 16 right-moving supercharges follow the same group-theoretic decomposition as the left-moving ones. E.g. for compactification on $T^8$,
one derives the following vertex operators for the  16  supercharges $\tilde Q_+^I$ and  $\tilde Q_-^{I'}$ 
 \begin{eqnarray}\label{supercharges2dleft}
\tilde Q_+^{  I} &=&
	\oint {d\bar z\over 2\pi i}e^{i w^{ I}_s\cdot \tilde H(\bar z)}e^{{\tilde\phi(\bar z)+i\tilde H_{0}(\bar z)\over 2}} \,,\nonumber\\
\tilde Q_-^{  I'} &=&
	\oint {d\bar z\over 2\pi i}e^{i w^{ I'}_c\cdot \tilde H(\bar z)}e^{{\tilde\phi(\bar z)-i\tilde H_{0}(\bar z)\over 2}} \, .
\end{eqnarray}
The two sets of supercharges correspond to the decomposition of the ten-dimensional spinor ${\bf{16}}_c^R$ to two dimensions in the same way as in eq.(\ref{decomp}).

For type IIA on $T^8$ however, the right-moving supercharges originate instead from the ${\bf{16}}_s$ spinor representation of $SO(10)$. In this case one obtains
 \begin{eqnarray}\label{supercharges2dlefta}
\tilde Q_+^{  I} &=&
	\oint {d\bar z\over 2\pi i}e^{i w^{ I}_c\cdot \tilde H(\bar z)}e^{{\tilde \phi(\bar z)+i \tilde H_{0}(\bar z)\over 2}} \,,\nonumber\\
\tilde Q_-^{  I'} &=&
	\oint {d\bar z\over 2\pi i}e^{i w^{ I'}_s\cdot \tilde H(\bar z)}e^{{\tilde \phi(\bar z)-i \tilde H_{0}(\bar z)\over 2}} \, ,
\end{eqnarray}
corresponding to the decomposition
\begin{equation}
{\bf{16}}_s^R \rightarrow (+\tfrac{1}{2},{\bf{8}}_c)^R+  (-\tfrac{1}{2},{\bf{8}}_s)^R\, .
\end{equation}

We see that in both cases for type IIA and type IIB on $T^8$ we have in total 32 supercharges in two dimensions  corresponding to the
$({\cal P},{\cal Q})=({\cal P}_L+{\cal P}_R,{\cal Q}_L+{\cal Q}_R)=(16,16)$ non-chiral
supersymmetry algebra, whereas  the heterotic string on  $T^8$ is characterized by a total of  16 supercharges realizing the $({\cal P}_L,{\cal Q}_L)=(8,8)$ non-chiral
supersymmetry algebra. In the next section we will discuss how the various orbifold projections act on the supercharges, which will result in several chiral, as well as non-chiral
superalgebras in two dimensions.

In general, the spectrum will fall into appropriate supermultiplets of the relevant $({\cal P},{\cal Q})$ supersymmetry algebra.
The generic supermultiplet is of course the supergravity multiplet, with $({\cal P},{\cal Q})$ supersymmetry.  However the graviton and all ${\cal P}+{\cal Q}$ gravitini are non-propagating in 2d. Therefore all supergravity multiplets in two dimensions only non-propagating states, which do
not contribute to the one-loop partition function of the theory. Notice, however, that even though the gravitino and spin-1/2 fermions in the gravity multiplet are non-propagating, they still contribute to gravitational anomalies \cite{AlvarezGaume:1983ig}.
Although there should  exist an off-shell formulation of the chiral $({\cal P},{\cal Q})$ supergravity action in two-dimensions, we will not display it here. 

We now turn to the matter fields, which arise from the compactification
from ten to two space-time dimensions. In addition to  left-
and right moving string excitations on the two-dimensional string
world-sheet, also the space-time spectrum is split into left- or
right moving states with respect to the two-dimensional target space.
In particular, massless fermions $\Psi_+$ or $\Psi_-$ can have
chirality $+1/2$ or $-1/2$ in two space-dimensions.  For non-chiral
theories with $({\cal P},{\cal P})$ space-time supersymmetry in two
dimensions, the left- and right-moving spectra match. However, for
chiral theories with $({\cal P},{\cal Q})$ supersymmetry
(${\cal P}\neq {\cal Q}$) the positive and negative chirality sectors are in general different
from each other. For example, in chiral theories with $({\cal P},0)$
supersymmetry the target-space left-movers are supersymmetric,
i.e. there is the same number of positive chirality fermions and
bosons, whereas the right-moving states are non-supersymmetric with
different numbers of negative chirality fermions and bosons.

In conclusion, the target-space left- and right-movers are associated
with two different sectors, which we will label by a parameter $\mu$:
the target space left-movers, i.e. the positive helicity states,
correspond to the choice $\mu=0$ and the right-movers with negative
helicity belong to $\mu=1$.  From the string point of view, the two
different cases $\mu=0,1$ are related to the two different
possibilities for the ``internal" part of the GSO projection on the
transverse string spectrum, as we shall discuss in the following
section.  While these are equivalent in dimensions $D=4k$, this is not
so in dimensions $D=4k+2$, where the supersymmetry algebra is chiral.

\section{Symmetric $T^8/\mathbb Z_4$ orbifold}
\label{sec:symmetric-Z4}

\subsection{Symmetric $T^8/\mathbb Z_4$ orbifold in type IIB with $(12,0)$ supersymmetry}\label{SymZ4}

Before dealing with the asymmetric orbifold constructions advertised in the introduction, it will be instructive to first revisit the symmetric $\mathbb Z_4$ case, involving the rotation of all eight coordinates of the internal 8-torus. Although these orbifolds have already been mentioned in \cite{Font:2004et}, the present analysis will serve as an opportunity to set up the notation and encounter some of the central ingredients that will be useful later on when we turn to the  asymmetric cases. In particular, we will consider IIA, IIB and Heterotic string theories compactified on $T^8/\mathbb Z_4$ orbifolds, with the orbifold action rotating all 8 internal super-coordinates as follows  
\begin{equation}
		\begin{array}{l c l}
			X^1 \to X^2  \quad &, &\quad X^2 \to - X^1 \,,\\
			X^3 \to -X^4  \quad &,&\quad X^4 \to  X^3 \,,\\
			X^5 \to X^6  \quad &, &\quad X^6 \to - X^5 \,,\\
			X^7 \to -X^8  \quad &,& \quad X^8 \to X^7 \,.\\
		\end{array}
	\label{orbDef}
\end{equation}
Although we display here only the $\mathbb Z_4$ action on the bosonic left and right moving coordinates $X^I$, worldsheet supersymmetry requires the exact same action also for the left and right moving worldsheet fermion superpartners, $\psi^I$. This orbifold action may be easily diagonalised by complexifying the corresponding super-coordinates.

In this section, we shall focus on type IIB case. Upon compactifying the 10d IIB string theory on the $T^8/\mathbb Z_4$ orbifold, one obtains a two dimensional theory with chiral supersymmetry. The full spectrum of the theory can be straightforwardly extracted from the one-loop partition function
\begin{equation}
	\begin{split}
Z= \frac{1}{4}\sum_{h,g\in\mathbb Z_4} &\left[\frac{1}{2}\sum_{a,b=0,1}(-1)^{a+b+\mu ab}\, \frac{\vartheta[^{a+h/2}_{b+g/2}]^2 \vartheta[^{a-h/2}_{b-g/2}]^2}{\eta^4}\, e^{-i\pi hg/2}\right] \,\frac{\Gamma^{\rm sym}_{8,8}[^h_g]}{\eta^8\bar\eta^8} \\
				\times  &\left[\frac{1}{2}\sum_{\bar a,\bar b=0,1}(-1)^{\bar a+\bar b+\mu \bar a\bar b}\, \frac{\bar\vartheta[^{\bar a+h/2}_{\bar b+g/2}]^2 \bar\vartheta[^{\bar a-h/2}_{\bar b-g/2}]^2}{\bar\eta^4}\, e^{+i\pi hg/2}\right]  \,.
	\end{split}\label{IIBsymPart}
\end{equation}
We denote by $\vartheta[^\alpha_\beta]\equiv\vartheta[^\alpha_\beta](0,\tau)$ the Jacobi theta constants, where we adopt the convention
\begin{equation}
	\vartheta[^\alpha_\beta](z,\tau) = \sum_{n\in\mathbb Z} q^{\frac{1}{2}\left(n-\alpha/2\right)^2}\,e^{2\pi i (z-\beta/2)(n-\alpha/2)} \,,
\end{equation}
in terms of the nome $q=e^{2\pi i \tau}$. They are holomorphic with respect to the complex structure parameter $\tau$ of the worldsheet torus $\Sigma_1$ and encode  the boundary conditions  $\alpha,\beta$ of worldsheet fermions $\psi^I$ along the $a$- and $b$- cycles of $\Sigma_1$, respectively. Similarly, the Dedekind $\eta$-functions are conventionally defined as $\eta(\tau)=\prod_{n>0}(1-q^n)$. The spin structures $a=0,1$ and $\bar a=0,1$ are associated to the left- and right- moving fermion numbers, respectively, while summing over $b,\bar b=0,1$ imposes the corresponding left- and right- moving GSO projections. Similarly, the orbifold sectors are labelled by $h=0,1,2,3$ while summing over $g=0,1,2,3$ enforces the $\mathbb Z_4$-invariance projection. The additional phases $e^{\mp i\pi hg/2}$, while cancelling each other in the total partition function, are inserted for later convenience in organising the various character blocks in a language that will be particularly useful once we turn to the asymmetric orbifold case. The contribution of the internal bosonic coordinates is encoded in the symmetrically twisted lattice $\Gamma_{8,8}^{\rm sym}[^h_g]$ carrying Lorentzian signature (8,8), and is defined as 
\begin{equation}
	\Gamma_{8,8}^{\rm sym}[^h_g] = \left\{\begin{array}{c l}
							2^8\sin^8\left(\frac{\pi}{4}\Lambda[^h_g]\right)\, \frac{\eta^{12} \bar\eta^{12}}{\left|\vartheta[^{1-h/2}_{1-g/2}]^2 \,\vartheta[^{1+h/2}_{1+g/2}]^2\right|^2} & \ \ ,\ \ (h,g)\neq(0,0) \\
							\Gamma_{8,8}(G,B) & \ \ ,\ \ (h,g)=(0,0) \\
							\end{array}\right. \,.
\end{equation}
Here $\Gamma_{8,8}(G,B)$ is the usual (untwisted) Narain lattice of $T^8$, as a function of the constant metric and antisymmetric tensor fields. The parameter $\Lambda[^h_g]$ is equal to $2$ for elements in the $\Gamma_0(2)$ orbit $[^0_2]$, $[^2_0]$, $[^2_2]$, and equal to $1$ otherwise.

Let us now comment on the particular choice of GSO phases $(-1)^{a+b+\mu ab}$ and $(-1)^{\bar a+\bar b+\mu\bar a\bar b}$ appearing in the left- and right- moving RNS fermion blocks. The factor $(-1)^{a+b+\bar a+\bar b}$ is standard, and present already in 10d. It is required by one-loop modular invariance and may be derived by imposing unitarity and cluster decomposition of higher loop amplitudes. It actually automatically implies the correct spin-statistics assignments. 

The additional phase factor $(-1)^{\mu(ab+\bar a\bar b)}$, with $\mu=0,1$ is more subtle and particular to twisted 2d constructions. In two dimensions, spacetime can be thought of as longitudinal, in the sense that it entirely spans the lightcone directions. States such as the graviton or the gravitini, which are constructed using string oscillators in spacetime directions are non-propagating and are, therefore, not captured by the simple one-loop partition function given above. This is because the longitudinal contribution $\vartheta[^a_b]/\eta$ is exactly cancelled by that of the (super)ghosts $\eta/\vartheta[^a_b]$ and a full covariant treatment \cite{Friedan:1985ey} is needed for the proper analysis of longitudinal states\footnote{Notice that it is actually possible to appropriately deform the longitudinal theta function contribution to the partition function by a Jacobi parameter $\vartheta[^a_b](0,\tau)\to\vartheta[^a_b](z,\tau)$ such that we can keep track of all states, including longitudinal ones. Although we have checked our results also using this method, we shall not employ it explicitly in this work, since it is more convenient to work with the lightcone partition function and revert to the full covariant formalism whenever needed.}. Consider, for instance, the left moving RNS sector. In the full covariant treatment and for a fixed superghost picture, the GSO projection is imposed at the full SO(10) level, and comes with an ambiguity concerning the choice of the SL(2) vacuum. Once this ambiguity is lifted by a choice of convention, the chirality of SO(10) spinors is completely fixed by the GSO projection. The lightcone partition function, on the other hand, can effectively see only the transverse SO(8) part of these spinors. Upon decomposing the $\textbf{16}$ spinorial representation under ${\rm SO}(10)\to {\rm SO}(2)\times {\rm SO}(8)$ we have, for instance, $\textbf{16}\to (+\frac{1}{2},\textbf{8}_s)\oplus(-\frac{1}{2},\textbf{8}_c)$, with SO(2) being the ``Euclideanised" Lorentz group in 2d. Specifically, the lightcone gauge treatment sets $X^+ = x^++p^+ t$ and $\psi^+=0$, with $t$ being the worldsheet proper time and, therefore, the partition function does not reproduce the full massless spectrum for chiral theories in 2d where the target space involves both left-moving and right-moving massless states. Massless propagating NS states ($a=0$) carry no SO(2) charge and a $\mu$-phase ambiguity does not arise. States in the R sector ($a=1$), however, do carry non-trivial SO(2) charge and the light cone gauge will select either the $\textbf{8}_s$ or the $\textbf{8}_c$ spinor, depending on the choice of $\mu=0,1$. For this reason, in our writing of the partition function \eqref{IIBsymPart}, we keep $\mu$ unspecified, with the understanding that a full analysis of the spectrum will require considering both the $\mu=0$ as well as the $\mu=1$ spectra, which will correspond to (propagating) states of positive and negative spacetime chirality, respectively.

The analysis of the full string spectrum in such $\mathbb Z_4$
orbifolds, and even more so in the asymmetric versions we shall be
considering in later sections, can be quite tedious and requires the
construction of SU(4) characters which are eigenmodes of the orbifold
action. Anticipating this, it is instructive and convenient to first
describe the massless sector of the theory from a geometric
perspective and then verify that these results are indeed reproduced
from the purely CFT orbifold analysis. The singularities in the
$T^8/\mathbb{Z}_4$ orbifold cannot be smoothed out in a Calabi-Yau
way, so there are no twisted sectors, and thus the cohomology
generators for the orbifold can be computed to be those of $T^8$ which
are left invariant by the orbifold action. Alternatively, the cohomology
may be computed from the knowledge of the internal CFT by exploiting
the isomorphism between the cohomology ring and the chiral ring of the
$\mathcal N=(2,2)$ worldsheet SCFT. The most straightforward way to do
this is by first computing the modified elliptic genus
\begin{equation}
	\hat Z(q,t,\bar q,\bar t) = {\rm Tr\, }_{RR}\left[ (-1)^f\, q^{L_0-c/24}\, \bar q^{\bar L_0-\bar c/24}\, t^{J_0}\, \bar t^{\bar J_0}\right] \,,
\end{equation}
where $f=J_0-\bar J_0$ in terms of the U(1) currents of the SCFT. From this, one may readily extract the Poincar\'e polynomial $P(t,\bar t) = (t\bar t)^{c/6} \hat Z(0,-t,0,-\bar t)$ of  the associated smooth manifold. Applying this to the $T^8/\mathbb Z_4$ orbifold at hand, we find the following Hodge square
\begin{equation}
	\left[
\begin{array}{ccccc}
 h_{0,0} & h_{0,1} & h_{0,2} & h_{0,3} & h_{0,4} \\
h_{1,0} & h_{1,1} & h_{1,2} & h_{1,3} & h_{1,4} \\
h_{2,0} & h_{2,1} & h_{2,2} & h_{2,3} & h_{2,4} \\
h_{3,0} & h_{3,1} & h_{3,2} & h_{3,3} & h_{3,4} \\
h_{4,0} & h_{4,1} & h_{4,2} & h_{4,3} & h_{4,4} \\
\end{array}
\right]=\left[
\begin{array}{ccccc}
 1 & 0 & 4 & 0 & 1 \\
 0 & 8 & 0 & 8 & 0 \\
 4 & 0 & 188 & 0 & 4 \\
 0 & 8 & 0 & 8 & 0 \\
 1 & 0 & 4 & 0 & 1
\end{array}
\right] \,,
\label{HodgeSquare}
\end{equation}
with Euler number $\chi(X)=240$. In terms of the Hodge numbers, one computes the number of self-dual and anti-self-dual 4-forms 
\begin{equation}
	\begin{split}
	b_4^{+} &= 2h_{1,3} +h_{1,1}-2h_{0,2}-1 = 15\,,\\
	b_4^{-}&=47+22h_{0,2}+4h_{1,3}+3h_{1,1}-2h_{1,2}=191 \,,
	\end{split}
\end{equation}
and also the Hirzebruch signature of the manifold
\begin{equation}
	\tau(X) = b_4^{-} - b_4^{+} =176 \,.
\end{equation}
A 10d M-W spinor $\textbf{16}_{c}$ of SO(10) decomposes under ${\rm SO}(2)\times {\rm SO}(8)$ as $(+\frac{1}{2},\textbf{8}_s) \oplus (-\frac{1}{2},\textbf{8}_c)$. On K\"ahler fourfolds, the SO(8) spinors can, in turn, be decomposed under the SU(4) holonomy in terms of gamma matrices acting on covariantly constant spinors, according to the group-theoretic decomposition
\begin{equation}
	\textbf{8}_s \to \textbf{1}+ \textbf{6} + \textbf{1} \quad,\quad \textbf{8}_c = \textbf{4}+\bar{\textbf{4}}\,.
\end{equation}
The coefficients of this decomposition correspond to $(0,q)$-forms. Hence, the singlets in the decomposition of $\textbf{8}_s$ are associated to $(0,0)$ and $(0,4)$ forms. To obtain the number of supersymmetries preserved by this compactification, one counts the zero modes of the Dirac operator arising from the above decomposition of the 10d gravitini. The latter correspond to those $(0,q)$ forms which are also harmonic. From $h_{0,0}=h_{0,4}=1$ and $h_{0,2}=4$ we see that there are 6 such zero modes. Since both 10d gravitini have the same SO(10) chirality, the reduction on the above fourfold gives rise to a 2d theory with chiral $(N,0)=(12,0)$ supersymmetry. More generally, for type IIB\footnote{In the type IIA case one finds instead that the theory has non-chiral $(N,N)$ supersymmetry with $N=2+h_{0,2}$.} on a fourfold with $h_{0,1}=h_{0,3}=0$ and $h_{0,2}\neq 0$, one finds $(N,0)$ chiral supersymmetry with $N=2(2+h_{0,2})$.

We may proceed with the analysis of the rest of the massless spectrum in a similar fashion. For brevity, we will mention here only the results which will, anyway, be re-derived below from the orbifold perspective. We have $4(h_{1,1}+h_{1,3}) = 64$ spacetime spin $1/2$ fields $\psi^+$ of positive chirality that will fit into chiral $(12,0)$ multiplets and $4h_{1,2}=0$ spinors of negative chirality which would have been supersymmetry singlets.
In other words, all massless fermions come with positive chirality. The supergravity multiplet is $(g_{\mu\nu},\phi,N\psi^-_\mu,N\psi^{+})$ and contains also the dilaton. Aside from the dilaton, there are $3h_{1,1}+2h_{1,3}+2h_{0,2}+1+b_4^{-}=240$ negative chirality scalars (susy singlets), and $3h_{1,1}+2h_{1,3}+2h_{0,2}+1+b_4^{+}=64$ positive chirality scalars. There are $64/16=4$ chiral matter multiplets of negative chirality $(16\phi^+,16\psi^+)$, since for $N=12$ supercharges, the massless matter multiplets actually have the same structure as in the $(16,0)$ case, even though the gravity multiplet does not. 

Since the theory is chiral, it is important to check  how the gravitational anomaly cancels in this case. The gravity multiplet contributes only through its fermionic states $(N\psi_\mu^-, N\psi^+)$, while chiral multiplets are of the form $(N'\phi^+,N'\psi^+)$. Here, $N=12$ which is the number of supercharges and $N'=16$ since the matter content of chiral multiplets for a $(12,0)$ theory in two spacetime dimensions is enlarged. The gravity multiplet contributes
\begin{equation}
	\frac{23N}{48}\,{\rm tr} R^2 +\frac{N}{48}\,{\rm tr} R^2 = \frac{N}{2}\,{\rm tr} R^2 \,,
\end{equation}
while the chiral multiplet contributes
\begin{equation}
	\frac{N'}{24}\,{\rm tr} R^2 + \frac{N'}{48}\,{\rm tr} R^2 = \frac{N'}{16}\,{\rm tr} R^2 \,.
\end{equation}
The 240 negative chirality scalars $\phi^-$ which are susy singlets contribute similarly $-240/24\,{\rm tr}R^2$ and there are no negative chirality fermions. Adding together these contributions one has
\begin{equation}
	\left(\frac{N}{2}+\frac{N'}{4}-\frac{240}{24} \right)\,{\rm tr}R^2 =0\,,
\end{equation}
which indeed vanishes for $N=12$ and $N'=16$, as desired. This is in accordance with the more general result in \cite{Dasgupta:1996yh}, in which the type IIB anomaly cancellation condition for a compactification on a fourfold may be expressed directly in terms of the number of supercharges $N$ and topological numbers of the fourfold
\begin{equation}
	24N-3\tau(X)+\chi(X)=0\,.
\end{equation}
The analogous condition for anomaly cancellation is much less straightforward in cases of asymmetric orbifolds, where no geometric interpretation is possible.

We are now ready to move on to the analysis of  the IIB theory in the orbifold theory. The CFT approach is exact to all orders in $\alpha'$ and enables one to access not only the massless spectrum, but also the full tower of massive excitations. Indeed, the vertex operator of any state in the theory may be straightforwardly constructed from the expansion of the partition function \eqref{IIBsymPart} in terms of characters. In the rest of the section, we set up the techniques necessary for deriving the character decomposition of the partition function and recover the same massless spectrum  as the one obtained above using the geometry of the fourfold. Since these techniques will set the basis for the analysis of the much less intuitive asymmetric orbifold constructions of subsequent sections, we will discuss them in some detail. 

In all orbifolds that we shall construct, all 8 RNS fermions spanning the transverse (internal) directions are twisted. By simple  redefinitions, it is actually convenient to group them all together, such that the relevant contribution to the partition function becomes
\begin{equation}
	\vartheta[^{a+h/2}_{b+g/2}]^2 \,\vartheta[^{a-h/2}_{b-g/2}]^2 = (-1)^{bh}\, \vartheta[^{a+h/2}_{b+g/2}]^4 \,.
\end{equation}
Out of these, we can construct characters with definite transformation under the internal GSO parity projection, associated to the transverse SO(8) charges,
\begin{equation}
	\begin{split}
	\Phi[^{a,h}_{+,g}] &= \frac{1}{2} \sum_{b=0,1} \frac{\vartheta[^{a+h/2}_{b+g/2}]^4}{\eta^4} \,(-1)^{bh}\,e^{-i\pi hg/2} \,, \\
	\Phi[^{a,h}_{-,g}] &= \frac{1}{2} \sum_{b=0,1} \frac{\vartheta[^{a+h/2}_{b+g/2}]^4}{\eta^4} \,(-1)^{bh+b}\,e^{-i\pi hg/2} \,, \\
	\end{split}
\end{equation}
but not under the orbifold action. In this way, $\Phi[^{a,h}_{+,g}]$ is invariant under GSO, while $\Phi[^{a,h}_{-,g}]$ picks a minus sign. This relation can be inverted if we identify $\xi=0$ with the $+$ (even) GSO parity and $\xi=1$ with the $-$ (odd) one as follows
\begin{equation}
	\frac{\vartheta[^{a+h/2}_{b+g/2}]^4}{\eta^4}\, (-1)^{bh} e^{-i\pi hg/2}= \sum_{\xi=0,1} (-1)^{b\xi} \,\Phi[^{a,h}_{\xi,g}] \,.
\end{equation}
We next define the characters with a definite eigenvalue under the $\mathbb Z_4$ action
\begin{equation}
	\chi[^{a,h}_{\pm, \lambda}] = \frac{1}{4}\sum_{g\in\mathbb Z_4} \Phi[^{a,h}_{\pm,g}]\,e^{2\pi i \lambda g/4}\,.
	\label{DefChar}
\end{equation}
This transforms by a phase $e^{2\pi i\lambda/4}$ under a $\mathbb Z_4$ action ($g\to g+1$). This relation may be inverted as follows
\begin{equation}
	\Phi[^{a,h}_{\pm,g}] = \sum_{\lambda\in \mathbb Z_4} \chi[^{a,h}_{\pm,\lambda}]\,e^{-2\pi i \lambda g/4} \,.
\end{equation}

By expressing the theta functions in the definition of $\chi$ in terms of their U(1) charges $Q_i=n_i-a/2-h/4$, where $n_i\in\mathbb Z$, we may explicitly sum over $b$ and $g$, \emph{i.e.} perform the GSO and $\mathbb Z_4$ projections, and derive  constraints on the corresponding charges. As before, we let $\xi=0$ denote the GSO even case, and  $\xi=1$ the GSO odd case. One then finds
\begin{equation}
	\begin{split}
				& n_1+n_2+n_3+n_4 = \xi \,{\rm mod}\,2  \,,\\
				& n_1+n_2+n_3+n_4 = (\lambda+2a) \,{\rm mod}\,4  \,.
	\end{split}
	\label{GSOZ4cond}
\end{equation}
These conditions must be satisfied simultaneously for states in $\chi$ which transform as $(-1)^\xi$ under GSO and as $e^{i\pi  \lambda/2}$  under $\mathbb Z_4$.
These states are associated with vertex operators of the form $e^{i Q\cdot H}$
\begin{equation}
	|Q\rangle = e^{i\sum_i(n_i-\frac{a}{2}-\frac{h}{4})H_i} |0\rangle\,,
\end{equation}
where $H_i(z)$ are the chiral bosons arising from the bosonization of the RNS fermions. Under the GSO transformation $H_i\to H_i + \pi$, this state transforms with a phase $e^{i\pi(\xi-h)}$. Similarly,  under the (redefined) $\mathbb Z_4$ action, $H_i\to H_i+\pi/2$, it acquires a phase $e^{i\pi(\lambda-h)/2}$.
Notice that the conditions \eqref{GSOZ4cond} imply $\xi = \lambda\, {\rm mod}\, 2$, regardless of whether the $\chi$ character is bosonic $a=0$ or fermionic $a=1$. Because of this, the transformation of the $\chi$ characters under GSO is essentially the square of the corresponding $\mathbb Z_4$ transformation, \emph{i.e.}
\begin{equation}
	\begin{split}
			{\rm GSO} \quad :& \quad H_i\to H_i+\pi \quad,\quad |Q\rangle \to e^{i\pi(\lambda-h)} \,|Q\rangle \,,\\
			\mathbb{Z}_4 \quad :& \quad H_i\to H_i+\tfrac{\pi}{2} \quad,\quad |Q\rangle \to e^{i\pi(\lambda-h)/2} \,|Q\rangle \,.\\
	\end{split}
\end{equation}

With these tools at our disposal, it is now possible to construct the characters $\chi$ for all orbifold sectors, as well as the corresponding vertex operators. The characters relevant for the purposes of the present work are collected in Appendix \ref{AppendixChar}.

For the extraction of the spectrum and the decomposition of the partition function in terms of characters, it will be convenient to define also the projected lattice characters, with definite eigenvalues under the $\mathbb Z_4$ orbifold. These eigenvalues will be parametrized again by $\lambda=0,1,2,3$, 
\begin{equation}
	\hat\Gamma^{\rm sym}[^h_\lambda] = \frac{1}{\eta^8 \bar\eta^8}\,\frac{1}{4}\sum_{g\in\mathbb Z_4}\Gamma_{8,8}^{\rm sym}[^h_g]\,e^{2\pi i \lambda g/4} \,.
\end{equation}	
The $q$-expansions of  the symmetrically twisted lattice characters can be found in Appendix \ref{AppendixLat}.


%
\subsubsection{Positive chirality spectrum ($\mu=0$ states)}
We will first describe the spectrum in the $\mu=0$ case. The first step is to decompose the partition function \eqref{IIBsymPart} in terms of the characters introduced above. For the generic sector $h=0,1,2,3$ we arrange the various contributions into NS-NS, R-R, NS-R and R-NS sectors as follows
\begin{equation}
	\begin{split}
	Z_{h}^{\rm NS-NS} &= \left( \chi[^{0,h}_{-,1}]\bar\chi[^{0,h}_{-,1}]+\chi[^{0,h}_{-,3}]\bar\chi[^{0,h}_{-,3}]\right) \hat\Gamma^{\rm sym}[^h_0]+\left(\chi[^{0,h}_{-,1}]\bar\chi[^{0,h}_{-,3}]+\chi[^{0,h}_{-,3}]\bar\chi[^{0,h}_{-,1}]\right)\hat\Gamma^{\rm sym}[^h_2] \\
	Z_{h}^{\rm R-R} &= \left( \chi[^{1,h}_{-,1}]\bar\chi[^{1,h}_{-,1}]+\chi[^{1,h}_{-,3}]\bar\chi[^{1,h}_{-,3}]\right) \hat\Gamma^{\rm sym}[^h_0]+\left(\chi[^{1,h}_{-,1}]\bar\chi[^{1,h}_{-,3}]+\chi[^{1,h}_{-,3}]\bar\chi[^{1,h}_{-,1}]\right)\hat\Gamma^{\rm sym}[^h_2] \\
	Z_{h}^{\rm NS-R} =& -\left( \chi[^{0,h}_{-,1}]\bar\chi[^{1,h}_{-,1}]+\chi[^{0,h}_{-,3}]\bar\chi[^{1,h}_{-,3}]\right) \hat\Gamma^{\rm sym}[^h_0]-\left(\chi[^{0,h}_{-,1}]\bar\chi[^{1,h}_{-,3}]+\chi[^{0,h}_{-,3}]\bar\chi[^{1,h}_{-,1}]\right)\hat\Gamma^{\rm sym}[^h_2] \\
	Z_{h}^{\rm R-NS} =& -\left( \chi[^{1,h}_{-,1}]\bar\chi[^{0,h}_{-,1}]+\chi[^{1,h}_{-,3}]\bar\chi[^{0,h}_{-,3}]\right) \hat\Gamma^{\rm sym}[^h_0]-\left(\chi[^{1,h}_{-,1}]\bar\chi[^{0,h}_{-,3}]+\chi[^{1,h}_{-,3}]\bar\chi[^{0,h}_{-,1}]\right)\hat\Gamma^{\rm sym}[^h_2] \\
	\end{split}
\end{equation}

This contains all information about the massless spectrum as well as the full tower of massive excitations. For our purposes, we will only extract the spectrum of massless propagating states. Using the explicit expansions in Appendices \ref{AppendixChar} and \ref{AppendixLat}, one may see that all twisted sectors are massive for $\mu=0$, and therefore, we can focus only on the untwisted one ($h=0$). We also need to remember that the right-moving characters $\bar\chi$ have opposite charges and transformation under $\mathbb Z_4$ with respect to the left-moving ones.  From the NS-NS sector we have 32 states
\begin{equation}
	\begin{split}
	\chi[^{0,0}_{-,1}]\bar\chi[^{0,0}_{-,1}] \quad :\quad \Psi^i \,\tilde{\bar\Psi}^j \quad,\quad \textbf{4}_{1}\times\bar{\textbf{4}}_{-1}\,,\\
	\chi[^{0,0}_{-,3}]\bar\chi[^{0,0}_{-,3}] \quad :\quad \bar\Psi^i \,\tilde\Psi^j \quad,\quad \bar{\textbf{4}}_{-1}\times \textbf{4}_{1} \,.\\
	\end{split}
\end{equation}
From the R-R sector we find
\begin{equation}
	\begin{split}
	\chi[^{1,0}_{-,1}]\bar\chi[^{1,0}_{-,1}] \quad:\quad \frac{1}{2}\left[\begin{array}{c c c c}
															- & + & + & + \\
															+ & - & + & + \\
															+ & + & - & + \\
															+ & + & + & - \\
														\end{array}\right]\times
											\frac{1}{2}\left[\begin{array}{c c c c}
															- & - & - & + \\
															- & - & + & - \\
															- & + & - & - \\
															+ & - & - & - \\
														\end{array}\right]\quad,\quad \bar{\textbf{4}}_{1}\times\textbf{4}_{-1}\,,\\
	\chi[^{1,0}_{-,3}]\bar\chi[^{1,0}_{-,3}] \quad:\quad \frac{1}{2}\left[\begin{array}{c c c c}
															- & - & - & + \\
															- & - & + & - \\
															- & + & - & - \\
															+ & - & - & - \\
														\end{array}\right]\times
											\frac{1}{2}\left[\begin{array}{c c c c}
															- & + & + & + \\
															+ & - & + & + \\
															+ & + & - & + \\
															+ & + & + & - \\
														\end{array}\right]\quad,\quad \textbf{4}_{-1}\times\bar{\textbf{4}}_{1}\,.														
	\end{split}													
\end{equation}
In this and subsequent expressions, whenever convenient, we adopt a notation according to which we use a matrix $M_{\alpha j}$ to denote the vertex operators $V_\alpha=\exp{(i M_{\alpha j} H^j)}$ associated to the Cartan weights of each state in the spinorial (or other) representations. In this way, the index $\alpha$ labels the states making up the representation, whereas $j=1,\ldots, 4$ spans the Cartan directions of SO(8).

From the NS-R sector, we obtain
\begin{equation}
	\begin{split}
			-\chi[^{0,0}_{-,1}]\bar\chi[^{1,0}_{-,1}] \quad:\quad \Psi^i  \times  \frac{1}{2}\left[\begin{array}{c c c c}
															- & - & - & + \\
															- & - & + & - \\
															- & + & - & - \\
															+ & - & - & - \\
														\end{array}\right]   \quad,\quad \textbf{4}_{1}\times\textbf{4}_{-1}   	\,, \\
			-\chi[^{0,0}_{-,3}]\bar\chi[^{1,0}_{-,3}] \quad:\quad \bar\Psi^i \times \frac{1}{2}\left[\begin{array}{c c c c}
															- & + & + & + \\
															+ & - & + & + \\
															+ & + & - & + \\
															+ & + & + & - \\
														\end{array}\right] \quad,\quad \bar{\textbf{4}}_{-1}\times\bar{\textbf{4}}_{1}\,,
	\end{split}
\end{equation}
Finally, the R-NS sector gives
\begin{equation}
	\begin{split}
			-\chi[^{1,0}_{-,1}]\bar\chi[^{0,0}_{-,1}] \quad:\quad    \frac{1}{2}\left[\begin{array}{c c c c}
															- & + & + & + \\
															+ & - & + & + \\
															+ & + & - & + \\
															+ & + & + & - \\
														\end{array}\right] \times \tilde{\bar\Psi}^i  \quad,\quad \bar{\textbf{4}}_{1}\times\bar{\textbf{4}}_{-1}   	\,, \\
			-\chi[^{1,0}_{-,3}]\bar\chi[^{0,0}_{-,3}] \quad:\quad  \frac{1}{2}\left[\begin{array}{c c c c}
															- & - & - & + \\
															- & - & + & - \\
															- & + & - & - \\
															+ & - & - & - \\
														\end{array}\right] \times \tilde\Psi^i\quad,\quad \textbf{4}_{-1}\times\textbf{4}_{1}\,,
	\end{split}
\end{equation}
These are indeed all the propagating massless states with positive spacetime chirality.

\subsubsection{Negative chirality spectrum ($\mu=1$ states)}
We now turn to the $\mu=1$ case, corresponding to states with negative spacetime chirality. For the generic sector $h=0,1,2,3$ we arrange the various contributions into NS-NS, R-R, NS-R and R-NS sectors as follows
\begin{equation}
	\begin{split}
	Z_{h}^{\rm NS-NS} &= \left( \chi[^{0,h}_{-,1}]\bar\chi[^{0,h}_{-,1}]+\chi[^{0,h}_{-,3}]\bar\chi[^{0,h}_{-,3}]\right) \hat\Gamma^{\rm sym}[^h_0]+\left(\chi[^{0,h}_{-,1}]\bar\chi[^{0,h}_{-,3}]+\chi[^{0,h}_{-,3}]\bar\chi[^{0,h}_{-,1}]\right)\hat\Gamma^{\rm sym}[^h_2] \\
	Z_{h}^{\rm R-R} &= \left( \chi[^{1,h}_{+,0}]\bar\chi[^{1,h}_{+,0}]+\chi[^{1,h}_{+,2}]\bar\chi[^{1,h}_{+,2}]\right) \hat\Gamma^{\rm sym}[^h_0]+\left(\chi[^{1,h}_{+,0}]\bar\chi[^{1,h}_{+,2}]+\chi[^{1,h}_{+,2}]\bar\chi[^{1,h}_{+,0}]\right)\hat\Gamma^{\rm sym}[^h_2] \\
	Z_{h}^{\rm NS-R} =& -\left( \chi[^{0,h}_{-,1}]\bar\chi[^{1,h}_{+,2}]+\chi[^{0,h}_{-,3}]\bar\chi[^{1,h}_{+,0}]\right) \hat\Gamma^{\rm sym}[^h_1]-\left(\chi[^{0,h}_{-,1}]\bar\chi[^{1,h}_{+,0}]+\chi[^{0,h}_{-,3}]\bar\chi[^{1,h}_{+,2}]\right)\hat\Gamma^{\rm sym}[^h_3] \\
	Z_{h}^{\rm R-NS} =& -\left( \chi[^{1,h}_{+,2}]\bar\chi[^{0,h}_{-,1}]+\chi[^{1,h}_{+,0}]\bar\chi[^{0,h}_{-,3}]\right) \hat\Gamma^{\rm sym}[^h_1]-\left(\chi[^{1,h}_{+,0}]\bar\chi[^{0,h}_{-,1}]+\chi[^{1,h}_{+,2}]\bar\chi[^{0,h}_{-,3}]\right)\hat\Gamma^{\rm sym}[^h_3] \\
	\end{split}
\end{equation}
 Consider first the untwisted sector $h=0$. The NS-NS sector is identical to the $\mu=0$ case. The R-R sector contains 40 massless states with SO(8) charges
\begin{equation}
	\begin{split}
	\chi[^{1,0}_{+,0}]\bar\chi[^{1,0}_{+,0}] \quad  &:\quad \frac{1}{2}\left[ \begin{array}{cccc}
															- & - & + & + \\
															- & + & - & + \\
															- & + & + & - \\
															+& -  & - & + \\
															+ &- & + & - \\
															+ & + & - & - \\
															\end{array}\right] \times
															\frac{1}{2}\left[ \begin{array}{cccc}
															- & - & + & + \\
															- & + & - & + \\
															- & + & + & - \\
															+& -  & - & + \\
															+ &- & + & - \\
															+ & + & - & - \\
															\end{array}\right]
															\quad,\quad 36 \,,\\
	\chi[^{1,0}_{+,2}]\bar\chi[^{1,0}_{+,2}] \quad &:\quad \frac{1}{2}\left[ \begin{array}{cccc}
															+ & + & + & + \\
															-  & -  & -  & - \\
															\end{array}\right] \times
												\frac{1}{2}\left[ \begin{array}{cccc}
															+ & + & + & + \\
															-  & -  & -  & - \\
															\end{array}\right]			
															\quad,\quad 4 \,.															
	\end{split}														
\end{equation}
The $h=1$ twisted sector contributes 16 massless states from the R-R sector
\begin{equation}
	\chi[^{1,1}_{+,2}]\bar\chi[^{1,1}_{+,2}]\,\hat\Gamma[^1_0] \quad:\quad \tfrac{1}{4}(+,+,+,+) \times \tfrac{1}{4}(-,-,-,-) \quad,\quad 16 \,.
\end{equation}
The $h=2$ twisted sector contributes 136 massless states from the R-R sector
\begin{equation}
	\chi[^{1,2}_{+,2}]\bar\chi[^{1,2}_{+,2}]\,\hat\Gamma[^2_0] \quad:\quad (0,0,0,0) \times (0,0,0,0) \quad,\quad 136 \,.
\end{equation}
Finally, the $h=3$ twisted sector contributes 16 massless states from the R-R sector
\begin{equation}
	\chi[^{1,3}_{+,2}]\bar\chi[^{1,3}_{+,2}]\,\hat\Gamma[^3_0] \quad:\quad \tfrac{1}{4}(-,-,-,-) \times \tfrac{1}{4}(+,+,+,+) \quad,\quad 16 \,.
\end{equation}

By taking into account all states (namely,  both chiralities $\mu=0$ and $\mu=1$) and computing the U(1) charges of R-R states with the definition $J=i\partial H_1+i\partial H_2-i\partial H_3-i\partial H_4$, as required due to the fusion $\vartheta[^{a+h/2}_{b+g/2}]^2\vartheta[^{a-h/2}_{b-g/2}]^2\to \vartheta[^{a+h/2}_{b+g/2}]^4$, one correctly reproduces the Hodge numbers given in \eqref{HodgeSquare}.

From the above spectrum of massless propagating fields, it is clear that only the states with positive spacetime chirality ($\mu=0$) are supersymmetric, whereas the negative chirality states are SUSY singlets. This is possible in two spacetime dimensions, where supersymmetry may be chiral. In what follows, we shall elaborate on the counting of  $(N,0)$ supercharges, and how the spectrum of massless propagating states organizes itself into (enhanced) chiral multiplets.

\subsubsection{Supersymmetry charges}
Let us consider first the left-moving (untwisted) sector. There are six $\mathbb Z_4$ invariant left-moving supercharges in the theory, which may be written in the $-1/2$ ghost picture as
\begin{equation}
	Q_{-1/2} = e^{-\phi/2-iH_0/2} \otimes \tfrac{1}{2}\left[ \begin{array}{c c c c}
												- & - & + & + \\
												- & + & - & + \\
												- & + & + & -\\
												+ & - & - & + \\
												+ & - & + & - \\
												+ & + & - & - \\
												\end{array}\right] \quad,\quad \textbf{6}_0\,.
\end{equation}
In accordance with the notation of the previous subsection, the six SO(8) charge vectors $Q_i$ simply stand for the vertex operators $e^{i Q_i H_i }$ with $i=1,\ldots 4$, which are left invariant under the $\mathbb Z_4$ action. They have GSO eigenvalue $G=+1$, according to the following  choice of GSO parity
\begin{equation}
	G = e^{i\pi ( -Q_\phi + Q_0 + Q_1+ Q_2+Q_3+Q_4 )}\,,
\end{equation}
which fixes the left-moving SL(2) vacuum convention. This is so, such that when the supercharges act on physical states in the spectrum, which themselves have $G=+1$ eigenvalue under the GSO operator, they produce superpartner states which are also physical states.

 Its BRST equivalent operator in the $+1/2$ picture reads
\begin{equation}
	Q_{+1/2} = e^{\phi/2+iH_0/2} \otimes \tfrac{1}{2}\left[ \begin{array}{c c c c}
												- & - & + & + \\
												- & + & - & + \\
												- & + & + & -\\
												+ & - & - & + \\
												+ & - & + & - \\
												+ & + & - & - \\
												\end{array}\right] \quad,\quad \textbf{6}_0\,.
\end{equation}
Similarly, there are 6 more invariant supercharges arising from the right-movers, which have the same spacetime chirality as the left-moving ones, i.e. $Q_{0}=\bar Q_0=-1/2$ in the $-1/2$ canonical superghost picture\footnote{Notice that, going from the $-1/2$ to the $+1/2$ ghost picture, the spacetime chirality is flipped. We shall adopt a convention according to which the spacetime chirality of states in the Ramond sector is always read in the $+1/2$ ghost picture.}. Therefore, the theory enjoys chiral $(N,0)=(12,0)$ supersymmetry. For the same reason, when $Q_{+1/2}$ acts on an NS-NS boson in the $-1$ ghost picture in order to create a fermion in the canonical $-1/2$ ghost picture, it will produce a spacetime fermion with positive chirality $\psi^+$.

Therefore, since the supersymmetry is chiral, aside from the gravity multiplet, only the states of positive chirality (described in the partition function by $\mu=0$) will be arranged into supersymmetry multiplets, while the negative chirality states will be SUSY singlets.

A convenient way of constructing the supersymmetry multiplets is to first identify the decomposition of the SO(8) characters $V_8, S_8, C_8$ under the R-symmetry ${\rm SO}(8)\to{\rm SU}(4)\times{\rm U}(1)$. One has
\begin{equation}
	\begin{split}
		&\textbf{8}_v \to \textbf{4}_1 + \textbf{4}_{-1} \,, \\
		&\textbf{8}_s \to \textbf{1}_2 + \textbf{1}_{-2} + \textbf{6}_0  \,, \\
		&\textbf{8}_c \to \textbf{4}_{-1}+\bar{\textbf{4}}_{1}  \,.
	\end{split}
\end{equation}
As usual, the subscript denotes the U(1) charge. In terms of components, the decomposition reads
\begin{equation}
	C_8 = \left[\begin{array}{c l}
					\frac{1}{2}\left[\begin{array}{c c c c}
							+ & + & + & - \\
							+ & + & - & + \\
							+ & - & + & + \\
							- & + & + & +\\
							\end{array}\right] & :\ \bar{\textbf{4}}_{1} \\
							& \\
					\frac{1}{2}\left[\begin{array}{c c c c}
							- & - & - & + \\
							- & - & + & - \\
							- & + & - & - \\
							+ & - & - & -\\
							\end{array}\right] & :\ \textbf{4}_{-1} \\
		\end{array} \right]\,,
\end{equation}
and similarly,
\begin{equation}
	S_8 = \left[\begin{array}{c l}
					\frac{1}{2}\left[\begin{array}{c c c c}
							+ & + & + & + \\
							\end{array}\right] & :\ \textbf{1}_{2} \\
							& \\
					\frac{1}{2}\left[\begin{array}{c c c c}
							- & - & - & - \\
							\end{array}\right] & :\ \textbf{1}_{-2} \\
							& \\
					\frac{1}{2}\left[\begin{array}{c c c c}
							+ & + & - & - \\
							+ & - & - & + \\
							+ & - & + & - \\
							- & - & + & +\\
							- & + & + & -\\
							- & + & - & +\\
							\end{array}\right] & :\ \textbf{6}_{0} \\
		\end{array} \right]\,,
\end{equation}
while, for the vector of SO(8) we have
\begin{equation}
	V_8 = \left[\begin{array}{c l}
					\left[\begin{array}{c c c c}
							1 &  &  &  \\
							 & 1 &  &  \\
							 &  & 1 &  \\
							 &  &  & 1\\
							\end{array}\right] & :\ \textbf{4}_{1} \\
							& \\
					\left[\begin{array}{c c c c}
							-1 &  &  &  \\
							 & -1 &  &  \\
							 &  & -1 &  \\
							 &  &  & -1\\
							\end{array}\right] & :\ \bar{\textbf{4}}_{-1} \\
		\end{array} \right]\,.
\end{equation}
Notice that the U(1) charge is also related to the $\mathbb Z_4$  charge via $Q_{\rm U(1)}/4 = Q_{\mathbb Z_4}$, with the eigenvalues of the states under the orbifold transformation being $e^{2\pi i Q_{\mathbb Z_4}}$.

We can now arrange the states of positive chirality ($\mu=0$) as follows
\begin{equation}
	\begin{split}
		{\rm NS-NS} \quad &:\quad \textbf{4}_1\times\bar{\textbf{4}}_{-1} + \bar{\textbf{4}}_{-1}\times\textbf{4}_1  \,, \\
		{\rm R-R} \quad & :\quad \textbf{4}_{-1}\times\bar{\textbf{4}}_{1} + \bar{\textbf{4}}_{1}\times\textbf{4}_{-1}  \,, \\
		{\rm NS-R} \quad & :\quad \textbf{4}_{1}\times\textbf{4}_{-1} + \bar{\textbf{4}}_{-1}\times \bar{\textbf{4}}_{1}  \,, \\
		{\rm R-NS} \quad & :\quad \textbf{4}_{-1}\times\textbf{4}_{1} + \bar{\textbf{4}}_{1}\times\bar{\textbf{4}}_{-1}  \,. \\
	\end{split}
\end{equation}
There are in total 64 massless bosons of positive chirality, and an equal number of fermions, of the same chirality, as expected by supersymmetry.
To see the action of the supersymmetry transformations, note the OPEs (or, fusion rule)
\begin{equation}
	\begin{split}
		&\textbf{6}_0 \cdot \textbf{4}_{-1} \to \bar{\textbf{4}}_{-1} \,, \\
		&\textbf{6}_0 \cdot \bar{\textbf{4}}_{1} \to \textbf{4}_{1}  \,,\\
	\end{split}
\end{equation}
and, clearly, also the inverse transformations exist.

For chiral $(N,0)$ supersymmetry in 2 dimensions, one would expect the massless states to be arranged into chiral multiplets $(N\phi^{+},N\psi^{+})$, where $\phi^{+}, \psi^{+}$ are  scalars and spin $1/2$ fermions of positive chirality, respectively. There are $n_{+}/N$ such multiplets, with $n_{+}$ being the total number of scalars of positive chirality. In the IIB theory we are analysing, $n_{+}=64$ and therefore, naively, $n_{+}/N= 16/3$, which is not an integer number. Actually, the structure of two dimensional  $N=12$ supersymmetry  is more subtle and the chiral matter multiplets become enlarged and arrange themselves as if they were multiplets of $N=16$. As we will explicitly see below,   there are indeed $64/16=4$ chiral multiplets in the theory.

To obtain the explicit multiplet structure, one needs to observe that NS-NS and R-R scalars are connected by 2 supersymmetry transformations (one from the left, and one from the right) and, hence, must lie in the same multiplet. This is achieved by taking linear combinations of the NS-NS and R-R states for the bosons and similarly NS-R and R-NS states for the fermions, as follows
\begin{equation}
	\begin{split}
			(16\phi^+,16\psi^+)^{(1)}=(\textbf{4}_1\times \bar{\textbf{4}}_{-1} + \bar{\textbf{4}}_{1}\times\textbf{4}_{-1} \ ,\ \bar{\textbf{4}}_{1}\times \bar{\textbf{4}}_{-1}+\textbf{4}_1\times\textbf{4}_{-1} ) \,,\\
			(16\phi^+,16\psi^+)^{(2)}=(\textbf{4}_1\times \bar{\textbf{4}}_{-1} - \bar{\textbf{4}}_{1}\times\textbf{4}_{-1} \ ,\ \bar{\textbf{4}}_{1}\times \bar{\textbf{4}}_{-1}-\textbf{4}_1\times\textbf{4}_{-1} ) \,,\\
			(16\phi^+,16\psi^+)^{(3)}=(\bar{\textbf{4}}_{-1}\times \textbf{4}_{1} + \textbf{4}_{-1}\times\bar{\textbf{4}}_{1} \ ,\ \textbf{4}_{-1}\times \textbf{4}_{1}+\bar{\textbf{4}}_{-1}\times\bar{\textbf{4}}_{1} ) \,,\\
			(16\phi^+,16\psi^+)^{(4)}=(\bar{\textbf{4}}_{-1}\times \textbf{4}_{1} - \textbf{4}_{-1}\times\bar{\textbf{4}}_{1} \ ,\ \textbf{4}_{-1}\times \textbf{4}_{1}-\bar{\textbf{4}}_{-1}\times\bar{\textbf{4}}_{1} ) \,.\\
	\end{split}
\end{equation} 
And indeed, we find 4 chiral multiplets containing 16 scalar and 16 fermion components each.



\subsection{Symmetric $T^8/\mathbb Z_4$ orbifold in type IIA with $(6,6)$ supersymmetry}

In this section we consider the compactification of type IIA string
theory on the same symmetric orbifold $T^8/\mathbb Z_4$, defined in
\eqref{orbDef}.\footnote{A specific limit in moduli space of this
  compactification, in the presence of D2 branes, gives rise to one of
  the $\mathcal{N}=3$ theories discussed in
  \cite{Garcia-Etxebarria:2015wns,Aharony:2016kai}.} Its partition
function is given by
\begin{equation}
	\begin{split}
Z= \frac{1}{4}\sum_{h,g\in\mathbb Z_4} &\left[\frac{1}{2}\sum_{a,b}(-1)^{a+b+\mu ab}\, \frac{\vartheta[^{a+h/2}_{b+g/2}]^2 \vartheta[^{a-h/2}_{b-g/2}]^2}{\eta^4}\, e^{-i\pi hg/2}\right] \,\frac{\Gamma^{\rm sym}_{8,8}[^h_g]}{\eta^8\bar\eta^8} \\
				\times  &\left[\frac{1}{2}\sum_{\bar a,\bar b}(-1)^{\bar a+\bar b+(\mu+1) \bar a\bar b}\, \frac{\bar\vartheta[^{\bar a+h/2}_{\bar b+g/2}]^2 \bar\vartheta[^{\bar a-h/2}_{\bar b-g/2}]^2}{\bar\eta^4}\, e^{+i\pi hg/2}\right]  \,.
	\end{split}
\end{equation}
The main difference between this partition function and the IIB case \eqref{IIBsymPart}, lies in the phase $(-1)^{(\mu+1)ab}$ phase, which ensures that the left- and right- moving spinors come with opposite Weyl chirality already at the 10d level. Contrary to the type IIB case, the IIA theory is not chiral and it is, therefore, sufficient to extract the physical spectrum by analysing only  the $\mu=0$ case. The internal manifold is, of course, identical to the one derived previously in the IIB case and, therefore, so are the elliptic genus and topological numbers ascribed to it.

The IIA theory enjoys $(6,6)$ supersymmetry. From the $\mu=0$ and $\mu=1$ cases, one obtains an equal number $2(h_{1,1}+h_{1,3}+h_{1,2})=32$ of positive and $32$ negative chirality Majorana-Weyl modulini. Similarly, there are $32+32$ supersymmetric partners, namely, chiral/anti-chiral scalars of positive/negative chirality (or, alternatively, 32 non-chiral scalars). These scalars arise as follows. There are $h_{1,1}+2h_{1,3}-2h_{0,2}=16$ real (non-chiral) scalars from the reduction of the metric, $h_{1,1}+2h_{0,2}=16$ from the NS-NS $B$-field, and $2h_{1,2}=0$ from the R-R 3-form. The R-R 1-form gives no dynamical fields. The dilaton goes into the gravity multiplet, together with 6 dilatini of positive and negative chirality, and together with the non-dynamical metric and gravitini. In two dimensional $(N,N)=(6,6)$ supersymmetry, the matter multiplets are the same as for $(8,8)$ and, therefore, there are $2(h_{1,2}+h_{1,3})/8=2$ chiral multiplets and $2h_{1,1}/8=2$ vector multiplets, but the content of the vector multiplets in 2d is the same as for chiral multiplets, plus a non-dynamical vector. Furthermore, type IIA theory in 2d generically has a B-field tadpole term  which, however, can be cancelled without introducing extra fluxes simply by adding $\chi/24=10$ fundamental strings \cite{Vafa:1995fj,Sen:1996na}. Each fundamental string has the matter content of its light-cone worldsheet fields, namely 8 real scalars and 8 positive and 8 negative chirality Majorana-Weyl fermions. In our case, this introduces $\chi/3N$ additional chiral multiplets. Since $N=6$ and the structure of chiral multiplets in this case is the same as for $N=8$, the addition of the fundamental strings corresponds to $\chi/24=10$ additional chiral multiplets.

\subsubsection{Positive chirality spectrum ($\mu=0$ states)}
We will first describe the spectrum in the $\mu=0$ case. For the generic sector $h=0,1,2,3$ we arrange the various contributions into NS-NS, R-R, NS-R and R-NS sectors as follows
\begin{equation}
	\begin{split}
	Z_{h}^{\rm NS-NS} &= \left( \chi[^{0,h}_{-,1}]\bar\chi[^{0,h}_{-,1}]+\chi[^{0,h}_{-,3}]\bar\chi[^{0,h}_{-,3}]\right) \hat\Gamma^{\rm sym}[^h_0]+\left(\chi[^{0,h}_{-,1}]\bar\chi[^{0,h}_{-,3}]+\chi[^{0,h}_{-,3}]\bar\chi[^{0,h}_{-,1}]\right)\hat\Gamma^{\rm sym}[^h_2] \\
	Z_{h}^{\rm R-R} &= \left( \chi[^{1,h}_{-,1}]\bar\chi[^{1,h}_{+,0}]+\chi[^{1,h}_{-,3}]\bar\chi[^{1,h}_{+,2}]\right) \hat\Gamma^{\rm sym}[^h_3]+\left(\chi[^{1,h}_{-,1}]\bar\chi[^{1,h}_{+,2}]+\chi[^{1,h}_{-,3}]\bar\chi[^{1,h}_{+,0}]\right)\hat\Gamma^{\rm sym}[^h_1] \\
	Z_{h}^{\rm NS-R} =& -\left( \chi[^{0,h}_{-,1}]\bar\chi[^{1,h}_{+,0}]+\chi[^{0,h}_{-,3}]\bar\chi[^{1,h}_{+,2}]\right) \hat\Gamma^{\rm sym}[^h_3]-\left(\chi[^{0,h}_{-,1}]\bar\chi[^{1,h}_{+,2}]+\chi[^{0,h}_{-,3}]\bar\chi[^{1,h}_{+,0}]\right)\hat\Gamma^{\rm sym}[^h_1] \\
	Z_{h}^{\rm R-NS} =& -\left( \chi[^{1,h}_{-,1}]\bar\chi[^{0,h}_{-,1}]+\chi[^{1,h}_{-,3}]\bar\chi[^{0,h}_{-,3}]\right) \hat\Gamma^{\rm sym}[^h_0]-\left(\chi[^{1,h}_{-,1}]\bar\chi[^{0,h}_{-,3}]+\chi[^{1,h}_{-,3}]\bar\chi[^{0,h}_{-,1}]\right)\hat\Gamma^{\rm sym}[^h_2] \\
	\end{split}
\end{equation}
The $\mu=1$ spectrum is the left-right mirror of this.  We now extract the massless states. In the untwisted $h=0$ sector, we have 32 NS-NS scalars of positive chirality
\begin{equation}
	\begin{split}
	\chi[^{0,0}_{-,1}]\bar\chi[^{0,0}_{-,1}] \quad :\quad \Psi^i \,\tilde{\bar\Psi}^j \quad,\quad \textbf{4}_{1}\times\bar{\textbf{4}}_{-1}\,,\\
	\chi[^{0,0}_{-,3}]\bar\chi[^{0,0}_{-,3}] \quad :\quad \bar\Psi^i \,\tilde\Psi^j \quad,\quad \bar{\textbf{4}}_{-1}\times \textbf{4}_{1} \,.\\
	\end{split}
\end{equation}
and an equal number of positive chirality R-NS fermions 
\begin{equation}
	\begin{split}
			-\chi[^{1,0}_{-,1}]\bar\chi[^{0,0}_{-,1}] \quad:\quad    \frac{1}{2}\left[\begin{array}{c c c c}
															- & + & + & + \\
															+ & - & + & + \\
															+ & + & - & + \\
															+ & + & + & - \\
														\end{array}\right] \times \tilde{\bar\Psi}^i  \quad,\quad \bar{\textbf{4}}_{1}\times\bar{\textbf{4}}_{-1}   	\,, \\
			-\chi[^{1,0}_{-,3}]\bar\chi[^{0,0}_{-,3}] \quad:\quad  \frac{1}{2}\left[\begin{array}{c c c c}
															- & - & - & + \\
															- & - & + & - \\
															- & + & - & - \\
															+ & - & - & - \\
														\end{array}\right] \times \tilde\Psi^i\quad,\quad \textbf{4}_{-1}\times\textbf{4}_{1}\,,
	\end{split}
\end{equation}
The twisted sectors are all massive.
\subsubsection{Negative chirality spectrum ($\mu=1$ states)}
The $\mu=1$ states are exactly the same as for the $\mu=0$ case but with left-right exchange and negative chirality. The NS-NS sector gives again 32 massless scalars of negative chirality
\begin{equation}
	\begin{split}
	\chi[^{0,0}_{-,1}]\bar\chi[^{0,0}_{-,1}] \quad :\quad \Psi^i \,\tilde{\bar\Psi}^j \quad,\quad \textbf{4}_{1}\times\bar{\textbf{4}}_{-1}\,,\\
	\chi[^{0,0}_{-,3}]\bar\chi[^{0,0}_{-,3}] \quad :\quad \bar\Psi^i \,\tilde\Psi^j \quad,\quad \bar{\textbf{4}}_{-1}\times \textbf{4}_{1} \,.\\
	\end{split}
\end{equation}
and an equal number of negative chirality NS-R fermions
\begin{equation}
	\begin{split}
			-\chi[^{0,0}_{-,1}]\bar\chi[^{1,0}_{-,1}] \quad:\quad \Psi^i  \times  \frac{1}{2}\left[\begin{array}{c c c c}
															- & - & - & + \\
															- & - & + & - \\
															- & + & - & - \\
															+ & - & - & - \\
														\end{array}\right]   \quad,\quad \textbf{4}_{1}\times\textbf{4}_{-1}   	\,, \\
			-\chi[^{0,0}_{-,3}]\bar\chi[^{1,0}_{-,3}] \quad:\quad \bar\Psi^i \times \frac{1}{2}\left[\begin{array}{c c c c}
															- & + & + & + \\
															+ & - & + & + \\
															+ & + & - & + \\
															+ & + & + & - \\
														\end{array}\right] \quad,\quad \bar{\textbf{4}}_{-1}\times\bar{\textbf{4}}_{1}\,,
	\end{split}
\end{equation}
Again, the twisted sectors are all massive.

The 6 $Q_{+1/2}$ supercharges of positive chirality act only on $\mu=0$ states, while the 6 $\tilde Q_{+1/2}$ supercharges of negative chirality act only on $\mu=1$ states, and the the theory has $(N,N)=(6,6)$ supersymmetry. The massless chiral multiplet has the structure $(N\phi^+, N\phi^-, N\psi^+, N\psi^-)$, and for $N=6$ as in our case, the multiplets become enlarged to the content of  $N=8$ ones. Thus, the perturbative spectrum gives a total of 4 chiral multiplets, in agreement with the previous analysis based on the smooth geometry.



\subsection{Symmetric $T^8/\mathbb Z_4$ orbifold in heterotic with $(6,0)$ supersymmetry}\label{HetSymSection}

We now consider heterotic ${\rm E}_8\times{\rm E}_8$ string theory compactified on the same symmetric orbifold $T^8/\mathbb Z_4$
\begin{equation}
	\begin{split}
Z= \frac{1}{4}\sum_{h,g\in\mathbb Z_4} &\left[\frac{1}{2}\sum_{a,b}(-1)^{a+b+\mu ab}\, \frac{\vartheta[^{a+h/2}_{b+g/2}]^2 \vartheta[^{a-h/2}_{b-g/2}]^2}{\eta^4}\, e^{-i\pi hg/2}\right] \,\frac{\Gamma^{\rm sym}_{8,8}[^h_g]}{\eta^8\bar\eta^8} \\
				\times  &\left[\frac{1}{2}\sum_{k,\ell}\,\frac{\bar\vartheta[^k_\ell]^4}{\bar\eta^4}\, \frac{\bar\vartheta[^{k+h/2}_{\ell+g/2}]^2 \bar\vartheta[^{k-h/2}_{\ell-g/2}]^2}{\bar\eta^4}\, e^{+i\pi hg/2}\right]\times\frac{1}{2}\sum_{\rho,\sigma}\frac{\bar\vartheta[^\rho_\sigma]^8}{\bar\eta^8}  \,.
	\end{split}
\end{equation}
As in the type IIB case, the chirality of the states is specified by the choice of $\mu=0,1$. Namely, the choice $\mu=0$ describes states of $+$ chirality, while the choice $\mu=1$ describes the states of opposite chirality. We will see below that the $\mu=0$ chirality states are supersymmetric, while the $\mu=1$ ones are not. Since the compactification manifold is the symmetric orbifold $T^8/\mathbb Z_4$ that we used before, the Hodge numbers are exactly the same as the ones we obtained in the type IIB case.

Moreover, this construction uses standard embedding and will have a non-vanishing tadpole or, equivalently, a non-vanishing coefficient in the corresponding anomaly polynomial. As in the type IIA case, the tadpole may be cancelled by introducing a certain number $n$ of fundamental strings, given by \cite{Vafa:1995fj,Sen:1996na}
\begin{equation}
	n = \frac{1}{8\pi} \int_{\mathcal F} d\mu\, A(\bar q,0,0)\,,
\end{equation}
where $\mu = d\tau_1 d\tau_2/\tau_2^2$ is the usual integration measure over the fundamental domain $\mathcal F$ of the worldsheet torus and  $A(\bar q,0,0)$ is simply the partition function of the internal CFT in the Ramond sector. Holomorphy and modularity force $A(\bar q,0,0)$ to take the form
\begin{equation}
	A(\bar q,0,0) = \alpha \bar J(\bar \tau)+\beta = \frac{\alpha}{\bar q}+\beta + \ldots \,,
\end{equation}
where $\alpha,\beta\in\mathbb Z$ and $J(\tau) = 1/q + \mathcal O(q)$ is essentially the usual invariant $j$-function with its constant term subtracted, i.e. $J(\tau)=j(\tau)-744$. After performing the modular integral, we obtain
\begin{equation}
	n = \frac{\beta}{24}-\alpha \,. 
	\label{HetFundString}
\end{equation}
Let us note that this matches also with the results in \cite{Font:2004et}
\begin{equation}
	n_{{\rm E}_8\times{\rm E}_8} = 15(2+h_{0,2}) \quad, \quad n_{{\rm SO}(32)} = 15(2+h_{0,2}) - \frac{\chi}{6} \,.
	\label{AMFundForm}
\end{equation}
which are, however, valid only in the case of standard embedding for geometric compactifications. Nevertheless, it should be noted that \eqref{HetFundString}, which is computed in terms of the index $A(\bar q,0,0)$ is more general and is, in particular, valid also for asymmetric orbifold constructions.

We can now evaluate this number in our heterotic symmetric $T^8/\mathbb Z_4$ orbifold. We have
\begin{equation}
	\begin{split}
			A(q,0,0)= &\frac{1}{4}\sum_{(h,g)\neq(0,0)} \frac{1}{2} \frac{\vartheta[^{1+h/2}_{1+g/2}]^2 \vartheta[^{1-h/2}_{1-g/2}]}{\eta^{12}}\, \frac{2^8 \sin^8(\frac{\pi}{4}\Lambda[^h_g])\,\eta^{12}\bar\eta^{12}}{\vartheta[^{1+h/2}_{1+g/2}]^2 \vartheta[^{1-h/2}_{1-g/2}] \bar\vartheta[^{1+h/2}_{1+g/2}]^2 \bar\vartheta[^{1-h/2}_{1-g/2}]} \\
			&\times \frac{1}{2} \sum_{k,\ell} \bar\vartheta[^k_\ell]^4 \bar\vartheta[^{k+h/2}_{\ell+g/2}]^2 \bar\vartheta[^{k-h/2}_{\ell-g/2}]^2 \, \frac{1}{2}\sum_{\rho,\sigma} \bar\vartheta[^\rho_\sigma]^8 \frac{1}{\bar\eta^{24}} \,.
	\end{split}
\end{equation}
As expected, the holomorphic $q-$part cancels out, and one is left with
\begin{equation}
	\begin{split}
	A(q,0,0)=& 2^3 \sum_{(h,g)\neq(0,0)} \sin^8\left(\tfrac{\pi}{4}\Lambda[^h_g]\right)\, \sum_{k,\ell} \frac{\bar\vartheta[^k_\ell]^4 \bar\vartheta[^{k+h/2}_{\ell+g/2}]^2 \bar\vartheta[^{k-h/2}_{\ell-g/2}]^2 }{\bar\eta^{12} \bar\vartheta[^{1-h/2}_{1-g/2}]^2 \bar\vartheta[^{1+h/2}_{1+g/2}]^2 } \sum_{\rho,\sigma} \bar\vartheta[^\rho_\sigma]^8 \\
		=& 3 j(\bar\tau) +2232 \,.
	\end{split}
\end{equation}
Hence, the number of fundamental heterotic strings that we must introduce in order to cancel the B-field tadpole is $n=2232/24-3=90$. We can also check this using \eqref{AMFundForm} which computes $n$ in terms of Hodge numbers, in perfect agreement with the result above.  In what follows, we will further check this number directly from the knowledge of the massless spectrum, by relating the tadpole coefficient to the anomaly polynomial of the 2d theory.

\subsubsection{Positive chirality spectrum ($\mu=0$ states)}
We begin with the positive chirality states. As usual, for the generic orbifold sector $h=0,1,2,3$, we arrange the various contributions into NS and R  contributions
\begin{equation}
	\begin{split}
	Z_h^{\rm NS}= \chi[^{0,h}_{-,1}]\Bigr[ &\hat\Gamma^{\rm sym}[^h_0](\bar V_8 \bar\chi[^{0,h}_{-,1}]+\bar C_8 \bar\chi[^{1,h}_{-,1}])+\hat\Gamma^{\rm sym}[^h_1](\bar O_8 \bar\chi[^{0,h}_{+,2}]+\bar S_8\bar\chi[^{1,h}_{+,2}])   \\
	  + &\hat\Gamma^{\rm sym}[^h_2](\bar V_8\bar\chi[^{0,h}_{-,3}]+\bar C_8\bar\chi[^{1,h}_{-,3}])+ \hat\Gamma^{\rm sym}[^h_3](\bar O_8\bar\chi[^{0,h}_{+,0}]+\bar S_8\bar\chi[^{1,h}_{+,0}]) \Bigr]  (\bar O_{16}+\bar S_{16}) \\
	  + \chi[^{0,h}_{-,3}] \Bigr[ &\hat\Gamma^{\rm sym}[^h_2](\bar V_8 \bar\chi[^{0,h}_{-,1}]+\bar C_8 \bar\chi[^{1,h}_{-,1}])+\hat\Gamma^{\rm sym}[^h_3](\bar O_8 \bar\chi[^{0,h}_{+,2}]+\bar S_8\bar\chi[^{1,h}_{+,2}])   \\
	  + &\hat\Gamma^{\rm sym}[^h_0](\bar V_8\bar\chi[^{0,h}_{-,3}]+\bar C_8\bar\chi[^{1,h}_{-,3}])+ \hat\Gamma^{\rm sym}[^h_1](\bar O_8\bar\chi[^{0,h}_{+,0}]+\bar S_8\bar\chi[^{1,h}_{+,0}]) \Bigr]  (\bar O_{16}+\bar S_{16}) \,. 
	\end{split}
\end{equation}
\begin{equation}
	\begin{split}
	Z_h^{\rm R}= -\chi[^{1,h}_{-,1}]\Bigr[ &\hat\Gamma^{\rm sym}[^h_0](\bar V_8 \bar\chi[^{0,h}_{-,1}]+\bar C_8 \bar\chi[^{1,h}_{-,1}])+\hat\Gamma^{\rm sym}[^h_1](\bar O_8 \bar\chi[^{0,h}_{+,2}]+\bar S_8\bar\chi[^{1,h}_{+,2}])   \\
	  + &\hat\Gamma^{\rm sym}[^h_2](\bar V_8\bar\chi[^{0,h}_{-,3}]+\bar C_8\bar\chi[^{1,h}_{-,3}])+ \hat\Gamma^{\rm sym}[^h_3](\bar O_8\bar\chi[^{0,h}_{+,0}]+\bar S_8\bar\chi[^{1,h}_{+,0}]) \Bigr]  (\bar O_{16}+\bar S_{16}) \\
	  - \chi[^{1,h}_{-,3}] \Bigr[ &\hat\Gamma^{\rm sym}[^h_2](\bar V_8 \bar\chi[^{0,h}_{-,1}]+\bar C_8 \bar\chi[^{1,h}_{-,1}])+\hat\Gamma^{\rm sym}[^h_3](\bar O_8 \bar\chi[^{0,h}_{+,2}]+\bar S_8\bar\chi[^{1,h}_{+,2}])   \\
	  + &\hat\Gamma^{\rm sym}[^h_0](\bar V_8\bar\chi[^{0,h}_{-,3}]+\bar C_8\bar\chi[^{1,h}_{-,3}])+ \hat\Gamma^{\rm sym}[^h_1](\bar O_8\bar\chi[^{0,h}_{+,0}]+\bar S_8\bar\chi[^{1,h}_{+,0}]) \Bigr]  (\bar O_{16}+\bar S_{16}) \,. 
	\end{split}
\end{equation}
Here only the untwisted sector $h=0$ gives rise to massless states. The first ${\rm E}_8$ gauge group factor breaks to ${\rm SO}(14)\times{\rm U}(1)$. The scalars are
\begin{equation}
	\begin{split}
	\chi[^{0,0}_{-,1}] (\bar V_8 \bar\chi[^{0,0}_{-,1}]+\bar C_8 \bar\chi[^{1,0}_{-,1}]) \quad :\quad 4\times (\textbf{64},\textbf{1}) \,,\\
	\chi[^{0,0}_{-,3}]\,\hat\Gamma^{\rm sym}[^0_3] \quad:\quad 16\times (\textbf{1},\textbf{1}) \,,
	\end{split}
\end{equation}
where by boldface we denote the representations of SO(14) and ${\rm E}_8$.
The fermions read
\begin{equation}
	\begin{split}
	-\chi[^{1,0}_{-,1}] (\bar V_8 \bar\chi[^{0,0}_{-,1}]+\bar C_8 \bar\chi[^{1,0}_{-,1}]) \quad :\quad 4\times (\textbf{64},\textbf{1}) \,,\\
	-\chi[^{1,0}_{-,3}]\,\hat\Gamma^{\rm sym}[^0_3] \quad:\quad 16\times (\textbf{1},\textbf{1}) \,.
	\end{split}
\end{equation}
In total, we have 544 massless scalars $\phi_+$ and 544 massless fermions $\psi_+$ of positive spacetime chirality, which organize themselves into 68 representations of $(6,0)$ chiral supersymmetry. As in previous cases, each massless chiral multiplet has the same content as an $(8,0)$ one.

\subsubsection{Negative chirality spectrum ($\mu=1$ states)}
We now turn to states of negative chirality, arising from $\mu=1$, and which are now supersymmetry singlets. For the generic sector $h=0,1,2,3$, once again, we arrange the various contributions into NS and R  contributions
\begin{equation}
	\begin{split}
	Z_h^{\rm NS}= \chi[^{0,h}_{-,1}]\Bigr[ &\hat\Gamma^{\rm sym}[^h_0](\bar V_8 \bar\chi[^{0,h}_{-,1}]+\bar C_8 \bar\chi[^{1,h}_{-,1}])+\hat\Gamma^{\rm sym}[^h_1](\bar O_8 \bar\chi[^{0,h}_{+,2}]+\bar S_8\bar\chi[^{1,h}_{+,2}])   \\
	  + &\hat\Gamma^{\rm sym}[^h_2](\bar V_8\bar\chi[^{0,h}_{-,3}]+\bar C_8\bar\chi[^{1,h}_{-,3}])+ \hat\Gamma^{\rm sym}[^h_3](\bar O_8\bar\chi[^{0,h}_{+,0}]+\bar S_8\bar\chi[^{1,h}_{+,0}]) \Bigr]  (\bar O_{16}+\bar S_{16}) \\
	  + \chi[^{0,h}_{-,3}] \Bigr[ &\hat\Gamma^{\rm sym}[^h_2](\bar V_8 \bar\chi[^{0,h}_{-,1}]+\bar C_8 \bar\chi[^{1,h}_{-,1}])+\hat\Gamma^{\rm sym}[^h_3](\bar O_8 \bar\chi[^{0,h}_{+,2}]+\bar S_8\bar\chi[^{1,h}_{+,2}])   \\
	  + &\hat\Gamma^{\rm sym}[^h_0](\bar V_8\bar\chi[^{0,h}_{-,3}]+\bar C_8\bar\chi[^{1,h}_{-,3}])+ \hat\Gamma^{\rm sym}[^h_1](\bar O_8\bar\chi[^{0,h}_{+,0}]+\bar S_8\bar\chi[^{1,h}_{+,0}]) \Bigr]  (\bar O_{16}+\bar S_{16}) \,. 
	\end{split}
\end{equation}
The NS states are of course identical as for the $\mu=0$ case and arise only from the untwisted sector. The difference comes in the fermions, which read
\begin{equation}
	\begin{split}
	Z_h^{\rm R}= -\chi[^{1,h}_{+,0}]\Bigr[ &\hat\Gamma^{\rm sym}[^h_1](\bar V_8 \bar\chi[^{0,h}_{-,1}]+\bar C_8 \bar\chi[^{1,h}_{-,1}])+\hat\Gamma^{\rm sym}[^h_2](\bar O_8 \bar\chi[^{0,h}_{+,2}]+\bar S_8\bar\chi[^{1,h}_{+,2}])   \\
	  + &\hat\Gamma^{\rm sym}[^h_3](\bar V_8\bar\chi[^{0,h}_{-,3}]+\bar C_8\bar\chi[^{1,h}_{-,3}])+ \hat\Gamma^{\rm sym}[^h_0](\bar O_8\bar\chi[^{0,h}_{+,0}]+\bar S_8\bar\chi[^{1,h}_{+,0}]) \Bigr]  (\bar O_{16}+\bar S_{16}) \\
	  - \chi[^{1,h}_{+,2}] \Bigr[ &\hat\Gamma^{\rm sym}[^h_3](\bar V_8 \bar\chi[^{0,h}_{-,1}]+\bar C_8 \bar\chi[^{1,h}_{-,1}])+\hat\Gamma^{\rm sym}[^h_0](\bar O_8 \bar\chi[^{0,h}_{+,2}]+\bar S_8\bar\chi[^{1,h}_{+,2}])   \\
	  + &\hat\Gamma^{\rm sym}[^h_1](\bar V_8\bar\chi[^{0,h}_{-,3}]+\bar C_8\bar\chi[^{1,h}_{-,3}])+ \hat\Gamma^{\rm sym}[^h_2](\bar O_8\bar\chi[^{0,h}_{+,0}]+\bar S_8\bar\chi[^{1,h}_{+,0}]) \Bigr]  (\bar O_{16}+\bar S_{16}) \,. 
	\end{split}
\end{equation}
For the fermions, there are contributions from all twisted and untwisted sectors. From the untwisted sector $h=0$ we have a total of 2096 massless fermionic states of negative chirality
\begin{equation}
	\begin{split}
		-\chi[^{1,0}_{+,0}] (\bar O_8 \bar\chi[^{0,0}_{+,0}]+\bar S_8 \bar\chi[^{1,0}_{+,0}])(\bar O_{16}+\bar S_{16}) \quad &:\quad  6 \times [(\textbf{91},\textbf{1})+(\textbf{1},\textbf{1})+(\textbf{1},\textbf{248}) ] \,,\\
		-\chi[^{1,0}_{+,2}] (\bar O_8 \bar\chi[^{0,0}_{+,2}]+\bar S_8 \bar\chi[^{1,0}_{+,2}])(\bar O_{16}+\bar S_{16}) \quad &:\quad 2\times [(\textbf{14},\textbf{1})+(\textbf{14},\textbf{1})] \,.
	\end{split}
\end{equation}
From the $h=1$ sector we have a total of 384 massless fermionic states of negative chirality
\begin{equation}
	\begin{split}
		-\chi[^{1,1}_{+,2}] \hat\Gamma^{\rm sym}[^1_0] (\bar O_8 \bar\chi[^{0,1}_{+,2}]+\bar S_8 \bar\chi[^{1,1}_{+,2}]) \quad &:\quad  16 \times (\textbf{14},\textbf{1})\,,\\
		-\chi[^{1,1}_{+,2}] \hat\Gamma^{\rm sym}[^1_2]\bar O_8 \bar\chi[^{0,1}_{+,0}] \quad &:\quad  160 \times (\textbf{1},\textbf{1})\,.\\
	\end{split}
\end{equation}
From the $h=2$ sector we have a total of 2144 massless fermionic states of negative chirality
\begin{equation}
	\begin{split}
		-\chi[^{1,2}_{+,2}] \hat\Gamma^{\rm sym}[^2_2] (\bar O_8 \bar\chi[^{0,2}_{+,0}]+\bar S_8 \bar\chi[^{1,2}_{+,0}]) \quad &:\quad  120 \times [(\textbf{1},\textbf{1})+ (\textbf{1},\textbf{1})]\,,\\
		-\chi[^{1,2}_{+,2}] \hat\Gamma^{\rm sym}[^2_0](\bar O_8 \bar\chi[^{0,2}_{+,2}]+\bar S_8\bar\chi[^{1,2}_{+,2}]) \quad &:\quad  136 \times (\textbf{14},\textbf{1})\,.\\
	\end{split}
\end{equation}
Finally, form the sector $h=3$ we have a total of 384 states of negative chirality
\begin{equation}
	\begin{split}
		-\chi[^{1,3}_{+,2}] \hat\Gamma^{\rm sym}[^3_0] (\bar O_8 \bar\chi[^{0,3}_{+,2}]+\bar S_8 \bar\chi[^{1,3}_{+,2}]) \quad &:\quad  16 \times (\textbf{14},\textbf{1})\,,\\
		-\chi[^{1,3}_{+,2}] \hat\Gamma^{\rm sym}[^3_2](\bar O_8 \bar\chi[^{0,3}_{+,0}]+\bar S_8\bar\chi[^{1,3}_{+,0}]) \quad &:\quad  160 \times (\textbf{1},\textbf{1})\,.\\
	\end{split}
\end{equation}

Adding all sectors together, we find a total of 5008 massless chiral fermions $\psi_{-}$ of negative chirality. We may now perform an independent check of the B-field tadpole coefficient $n=90$ that was computed earlier, by exploiting the fact that the tadpole arises from 10d anomaly canceling term $\int B\wedge X_8$, where $X_8$ is a particular curvature and field strength 8-form. In 2d, the anomaly would be cancelled by the term $n\int B$ in the Green-Schwarz mechanism, with $n=\frac{1}{48(2\pi)^4} \int X_8$. 

The anomaly coefficient in the 2d theory with $(N,0)$ supersymmetry is obtained as 
\begin{equation}
	-n=\frac{N}{2} + n_{+} \frac{N'}{16} + n_\phi^{-} \left(-\frac{1}{24}\right) + n_{\psi}^{-} \left(-\frac{1}{48}\right) \,.
\end{equation}
The contribution $N/2$ is the contribution of the gravity multiplet,  $N'/16$ is that of a chiral multiplet (positive chirality) which contains $N'$ chiral scalars, $-1/24$ is the contribution of a scalar of negative chirality and $-1/48$ is the contribution of a fermion of negative chirality. In our case $N=6$ but the chiral multiplets actually have $N'=8$. There are  $n_{+}=544/8 = 68$ chiral multiplets, and we have $544$ negative chirality bosons and $5008$ negative chirality fermions. Putting everything together we verify indeed $n=90$  in accordance with the previous calculation. The tadpole can be cancelled by introducing $n=90$ elementary heterotic strings, which amounts to 90 additional chiral multiplets in the spectrum of the theory.


\section{Asymmetric $T^8/\mathbb Z_4$ orbifold}
\label{sec:asymmetric-Z4}

\subsection{Asymmetric $T^8/\mathbb Z_4$ orbifold in type IIB with $(24,8)$ supersymmetry}\label{SecAsymIIB}

We are now ready to consider asymmetric $T^8/\mathbb Z_4$ orbifold constructions, where the left-moving coordinates are rotated according to the $\mathbb Z_4$ action \eqref{orbDef}, whereas the right-moving ones are inert. Such an orbifold is heavily left-right asymmetric and exhibits strong similarities to the worldsheet realisation of special non-geometric fluxes considered in \cite{CFL,CFKL}. These are inherently stringy constructions without geometric duals or, equivalently, with the $\mathbb Z_4$ action on the super-coordinates lying outside the conjugacy class of geometric T-duality actions. As a result, such constructions cannot be cast into a geometric duality frame and are typically constrained to live only at special points in moduli space. 

In essence, the asymmetry of the construction implies  the geometric moduli are now charged under the $\mathbb Z_4$ action. If  the $\mathbb Z_4$ rotation were also coupled to a translation to render the action free, which from the point of view of target space would correspond to turning on a non-geometric flux background,  a non-trivial scalar potential would be generated for these moduli at tree level, and the orbifold theory would live at the minimum of this potential with vanishing vacuum energy \cite{CFL,CFKL}. In the case at hand, the corresponding orbifold action has fixed points and the non-invariant moduli are in fact projected out of the physical spectrum.

In general, asymmetric orbifold constructions are not automatic but are heavily constrained by modular invariance, which implies unitarity at all genera and the correct particle interpretation of the theory. In general, such constructions are only possible at special points of enhanced symmetry in moduli space that ensure the appropriate factorisation of the underlying left- and right- moving CFT on which the asymmetric orbifold acts.

We will realise our construction at the fermionic point in $T^8$ moduli space, where the torus factorises into a product of eight circles at the special radii $R_I=1/\sqrt{2}$, so that we can consistently fermionise all coordinates as follows
\begin{align}
	\partial X^I = y^I w^I  \,,
\end{align}
and similarly for the right movers. Here, $y^I(z)$ and $w^I(z)$ are auxiliary left-moving fermions on the worldsheet, that are introduced to replace the CFT of a left-moving boson at $R=1/\sqrt{2}$ with that of a pair of real left-moving fermions. As a result of sitting at this special point, it is possible to obtain a representation of the (untwisted) Narain lattice $\Gamma_{8,8}$ entirely in terms of theta functions
\begin{align}
	\Gamma_{8,8} = \frac{1}{2}\sum_{\gamma,\delta=0,1} \left| \vartheta\bigr[^\gamma_\delta\bigr] \right|^{16} \,.
\end{align}
We are now ready to introduce the $\mathbb Z_4$ action at the level of the partition function. Since all bosonic $T^8$ coordinates have been fermionised, the asymmetric $\mathbb Z_4$ action will be introduced entirely at the level of the left-moving worldsheet fermions $\{ \psi^I, y^I, w^I\}$. It is straightforward to see that the action
\begin{align}
	\begin{split}
	\begin{array}{l l l}
	\psi^1 \to \psi^2  \quad,&  y^1\to y^2  \quad,& w^1\to w^2\,, \\
	\psi^2 \to -\psi^1  \quad,&  y^2\to -y^1 \quad,& w^2\to w^1\,, \\
	\psi^3 \to -\psi^4  \quad,&  y^3\to -y^4  \quad,& w^3\to w^4\,, \\
	\psi^4 \to \psi^3  \quad,&  y^4\to y^3 \quad,& w^4\to w^3\,, \\
	\psi^5 \to \psi^6  \quad,&  y^5\to y^6  \quad,& w^5\to w^6\,, \\
	\psi^6 \to -\psi^5  \quad,&  y^6\to -y^5 \quad,& w^6\to w^5\,, \\
	\psi^7 \to -\psi^8  \quad,&  y^7\to -y^8  \quad,& w^7\to w^8\,, \\
	\psi^8 \to \psi^7  \quad,&  y^8\to y^7 \quad,& w^8\to w^7\,, \\
	\end{array}
	\end{split}
\label{Z4orbaction}
\end{align}
correctly induces the left-moving  part of the $\mathbb Z_4$ transformation in \eqref{orbDef}. It constitutes a well defined chiral action on the left-moving d.o.f. under which the internal part of the worldsheet super-current
\begin{equation}
	T_F(z) \sim \sum_{I=1}^8 \psi^I y^I w^I \,,
\end{equation}
remains invariant. Diagonalising the action \eqref{Z4orbaction} in terms of complexified fermions allows one to read the boundary conditions of each, and hence arrive at the one-loop partition function of the 2d type IIB theory on $T^8/\mathbb Z_4$
\begin{equation}
	\begin{split}
Z= \frac{1}{4}\sum_{h,g\in\mathbb Z_4} &\left[\frac{1}{2}\sum_{a,b}(-1)^{a+b+\mu ab}\, \frac{\vartheta[^{a+h/2}_{b+g/2}]^2 \vartheta[^{a-h/2}_{b-g/2}]^2}{\eta^4}\, e^{-i\pi hg/2}\right] \,\frac{\Gamma^{\rm asym}_{8,8}[^h_g]}{\eta^8\bar\eta^8}\,e^{-i\pi hg/2} \\
				\times  &\left[\frac{1}{2}\sum_{\bar a,\bar b}(-1)^{\bar a+\bar b+\mu \bar a\bar b}\, \frac{\bar\vartheta[^{\bar a}_{\bar b}]^4}{\bar\eta^4}\, \right]  \,.
	\end{split}
\end{equation}
where the asymmetrically twisted (8,8) lattice is now also represented in terms of Jacobi thetas
\begin{equation}
	\Gamma_{8,8}^{\rm asym}\bigr[^h_g\bigr] = \frac{1}{2}\sum_{\gamma,\delta=0,1} \vartheta\bigr[^\gamma_\delta\bigr]^2\,\vartheta\bigr[^{\gamma+h/2}_{\delta+g/2}\bigr]^2\,\vartheta\bigr[^{\gamma-h/2}_{\delta-g/2}\bigr]^2\,\vartheta\bigr[^{\gamma-h}_{\delta-g}\bigr]\,\vartheta\bigr[^{\gamma+h}_{\delta+g}\bigr]\, \bar\vartheta\bigr[^\gamma_\delta\bigr]^8 \,.
\end{equation}
In some cases, the symmetrically twisted Narain lattice may be further simplified by using  the following generalization of Jacobi's triple product identity
\begin{equation}
	\vartheta[^\gamma_\delta]^2\,\vartheta[^{\gamma+h/2}_{\delta+g/2}]^2\,\vartheta[^{\gamma-h/2}_{\delta-g/2}]^2\,\vartheta[^{\gamma-h}_{\delta-g}]\,\vartheta[^{\gamma+h}_{\delta+g}] = 2^4 (-1)^{hg} \,\sin^4\left(\frac{\pi}{4}\Lambda[^h_g]\right)\, \frac{\eta^{12}}{\vartheta[^{1-h/2}_{1-g/2}]^2\,\vartheta[^{1+h/2}_{1+g/2}]^2} \,.
\end{equation}
This identity is valid for $(h,g)\neq(0,0)$ and whenever the l.h.s. does not vanish, i.e. $(\gamma,\delta)\neq(1,1)$, $(\gamma+h/2,\delta+g/2)\neq(1,1){\rm mod}\,2$ and $(\gamma+h,\delta+g)\neq(1,1){\rm mod}\,2$. Similarly to the symmetric case, we also define the asymmetric lattices with definite eigenvalue with respect to the orbifold 
\begin{equation}
	\hat\Gamma^{\rm asym}[^h_\lambda] = \frac{1}{\eta^8 \bar\eta^8}\,\frac{1}{4}\sum_{g\in\mathbb Z_4}\Gamma_{8,8}^{\rm asym}[^h_g]\,e^{2\pi i \lambda g/4} e^{-i\pi hg/2}\,.
\end{equation}	
Notice that we added the additional phase $e^{-ihg/2}$ which is required for modular invariance. The explicit expansions of these asymmetrically twisted lattice characters are assembled in Appendix \ref{AppendixLatAsym}.

We will see in what follows that type IIB theory compactified on this asymmetric orbifold enjoys chiral $(24,8)$ supersymmetry. Moreover, in the absence of a geometric background, the vanishing of gravitational anomalies should be checked by explicitly deriving the massless spectrum. As in the previous symmetric cases, the spacetime helicity of the states is specified by the choice of $\mu=0,1$. Namely, the choice $\mu=0$ describes states of $+$ spacetime chirality, while the choice $\mu=1$ describes the states of opposite chirality. We will show below that all propagating positive helicity states are massive, and there are 16 chiral multiplets with negative helicity arising from the $\mu=1$ sector.

\subsubsection{Positive chirality spectrum ($\mu=0$ states)}
The analysis of the spectrum works in a similar fashion, as in the symmetric case. We begin with the positive chirality states, $\mu=0$, and for the generic sector $h=0,1,2,3$ we obtain the expansion of the various contributions into NS-NS, R-R, NS-R and R-NS blocks 
\begin{equation}
	\begin{split}
	Z_{h}^{\rm NS-NS} &= \chi[^{0,h}_{-,1}]\,\hat\Gamma^{\rm asym}[^h_3]\,\bar V_8+\chi[^{0,h}_{-,3}]\,\hat\Gamma^{\rm asym}[^h_1]\,\bar V_8 \\
		Z_{h}^{\rm R-R} &= \chi[^{1,h}_{-,1}]\,\hat\Gamma^{\rm asym}[^h_3]\,\bar C_8+\chi[^{1,h}_{-,3}]\,\hat\Gamma^{\rm asym}[^h_1]\,\bar C_8 \\
	Z_{h}^{\rm NS-R} &= -\chi[^{0,h}_{-,1}]\,\hat\Gamma^{\rm asym}[^h_3]\,\bar C_8-\chi[^{0,h}_{-,3}]\,\hat\Gamma^{\rm asym}[^h_1]\,\bar C_8 \\
	Z_{h}^{\rm R-NS} &= -\chi[^{1,h}_{-,1}]\,\hat\Gamma^{\rm asym}[^h_3]\,\bar V_8-\chi[^{1,h}_{-,3}]\,\hat\Gamma^{\rm asym}[^h_1]\,\bar V_8 \\
	\end{split}
\end{equation}
Using the results in Appendices \ref{AppendixChar} and \ref{AppendixLatAsym}, it is straightforward to check that there are no propagating massless states of positive spacetime chirality in any sector $h=0,1,2,3$.

\subsubsection{Negative chirality spectrum ($\mu=1$ states)}
Moving on to the $\mu=1$ case, we perform the analogous  expansion into NS-NS, R-R, NS-R and R-NS blocks valid for the generic sector $h=0,1,2,3$  
\begin{equation}
	\begin{split}
	Z_{h}^{\rm NS-NS} &= \chi[^{0,h}_{-,1}]\,\hat\Gamma^{\rm asym}[^h_3]\,\bar V_8+\chi[^{0,h}_{-,3}]\,\hat\Gamma^{\rm asym}[^h_1]\,\bar V_8 \\
		Z_{h}^{\rm R-R} &= \chi[^{1,h}_{+,0}]\,\hat\Gamma^{\rm asym}[^h_0]\,\bar S_8+\chi[^{1,h}_{+,2}]\,\hat\Gamma^{\rm asym}[^h_2]\,\bar S_8 \\
	Z_{h}^{\rm NS-R} &= -\chi[^{0,h}_{-,1}]\,\hat\Gamma^{\rm asym}[^h_3]\,\bar S_8-\chi[^{0,h}_{-,3}]\,\hat\Gamma^{\rm asym}[^h_1]\,\bar S_8 \\
	Z_{h}^{\rm R-NS} &= -\chi[^{1,h}_{+,0}]\,\hat\Gamma^{\rm asym}[^h_0]\,\bar V_8-\chi[^{1,h}_{+,2}]\,\hat\Gamma^{\rm asym}[^h_2]\,\bar V_8 \\
	\end{split}
\end{equation}
Massless states with negative spacetime helicity arise only from $a=0$. In the untwisted $h=0$ sector we have 48 negative chirality fermions and an equal number of negative helicity RR scalars
\begin{equation}
	-\chi[^{1,0}_{+,0}]\,(\bar V_8-\bar S_8) \quad:\quad \textbf{6}\times(\bar V_8 - \bar S_8)\,,
\end{equation}
where $\textbf{6}$ is the antisymmetric representation of SU(4). In the $h=1$ sector we find 16 negative chirality fermions and an equal number of negative chirality RR scalars
\begin{equation}
	-\chi[^{1,1}_{+,2}]\,\hat\Gamma^{\rm asym}[^1_2]\,(\bar V_8-\bar S_8) \quad:\quad \tfrac{1}{4}[++++]\cdot(2q^{3/8})\cdot(\bar V_8-\bar S_8)\,.
\end{equation}
The notation $(2q^{3/8})$ within brackets denotes the 2 chiral operators of conformal weight $(\frac{3}{8},0)$ arising from the twisted lattice. Whenever convenient, we shall utilise this shorthand notation in what follows to refer to contributions involving the twisted lattice. In the $h=2$ twisted sector we find 48 negative chirality fermions and similarly for the RR scalars
\begin{equation}
	-\chi[^{1,2}_{+,2}]\,\hat\Gamma^{\rm asym}[^2_2]\,(\bar V_8-\bar S_8) \quad:\quad 1\cdot(6q^{1/2})\cdot(\bar V_8-\bar S_8)\,.
\end{equation}
Finally, in the $h=3$ sector we have 16 massless fermions of negative chirality and an equal number of RR scalars
\begin{equation}
	-\chi[^{1,3}_{+,2}]\,\hat\Gamma^{\rm asym}[^3_2]\,(\bar V_8-\bar S_8) \quad:\quad \tfrac{1}{4}[----]\cdot(2q^{3/8})\cdot(\bar V_8-\bar S_8) \,.
\end{equation}

\subsubsection{Supersymmetry charges}

In covariant formalism, we may now construct the vertex operators associated to the spacetime supercharges of the theory. In IIB theory we conventionally choose the GSO parity $G=+1$ to be
\begin{equation}
	G = e^{i\pi ( -Q_\phi + Q_0 + Q_1+ Q_2+Q_3+Q_4 )}\,,
\end{equation}
for the left-movers, and make an identical choice for the right-movers
\begin{equation}
	\tilde G = e^{i\pi ( -\bar Q_\phi + \bar Q_0 + \bar Q_1+ \bar Q_2+ \bar Q_3+ \bar Q_4 )}\,.
\end{equation}

We conventionally choose to display all supercharges in the $+1/2$ super-ghost picture. From the worldsheet left-movers and the untwisted sector $h=0$, we have 6 invariant supercharges with positive spacetime helicity ($Q_0=+1/2$)
\begin{equation}
	Q_{+1/2} = e^{\phi/2+iH_0/2} \otimes \tfrac{1}{2}\left[ \begin{array}{c c c c}
												- & - & + & + \\
												- & + & - & + \\
												- & + & + & -\\
												+ & - & - & + \\
												+ & - & + & - \\
												+ & + & - & - \\
												\end{array}\right] \quad,\quad \textbf{6}_0\,.
\end{equation}
These 6 invariant operators are exactly the same as the ones found in the untwisted sector of either the left-  or the right- moving sectors of the symmetric orbifolds discussed in previous sections. In the present asymmetric construction, however, this is not the end of the story. Indeed, if we assume that that the left-movers only contribute the above 6 positive chirality operators then, together  with the 8 positive and 8 negative chirality supercharges from the right-movers, one would obtain a $(14,8)$ theory. The gravity multiplet would then contribute $14\cdot\frac{1}{2} -8\cdot\frac{1}{2} = 3$ to the anomaly. The propagating $\mu=0$ states are all massive, but from the $\mu=1$ states one finds a total of 128 massless scalars and 128 massless fermions of negative chirality, contributing $-128\cdot\frac{1}{24}-128\cdot\frac{1}{48}=-8$ to the gravitational anomaly, which would then fail to cancel against the gravity multiplet. One, therefore, arrives to the na\"ive (and necessarily incorrect \cite{Schellekens:1986yi,Lerche:1987qk}) conclusion that the theory is anomalous, despite being modular invariant.

In fact, what this na\"ive anomaly counting really indicates is that there exist additional degrees of freedom in the theory that we have missed. This is not very surprising since in 2d, as already mentioned in previous sections, the fermionic content of the gravity multiplet is built out of string oscillators in the longitudinal directions and the lightcone partition function is blind to such states. It is then natural to ask where such additional supercharges could come from. Because the orbifold is fully asymmetric and forced to live at a special point in moduli space, it is possible for the twisted R-NS sectors to provide additional $(\frac{1}{2},0)$ conformal operators which, in turn, can be sewed together with appropriate string oscillators from the right-movers in order to construct  additional supercharges (or gravitini).

These can be easily extracted from the vertex operators of massless fermions of the previous section simply by replacing the right moving NS character $\bar V_8$ with the right moving worldsheet fermion oscillator $\tilde\psi^\mu$ in the spacetime directions, an operation that preserves the right-moving GSO parity. These are states that involve string oscillators in the lightcone directions and, therefore, do not propagate nor do they appear in the lightcone partition function, as explained earlier. Nevertheless, they still do contribute to the construction of the gravity multiplet and, hence, to the gravitational anomaly.   Explicitly, for $h=1,2,3$, they may be represented as
\begin{equation}
	\begin{split}
			e^{\phi/2+iH_0/2}\otimes\left(-\chi[^{1,h}_{+,2}]\,\hat\Gamma^{\rm asym}[^h_2]\right)\otimes e^{-\tilde\phi}\,\tilde\psi^\mu  \quad ,\quad \begin{array}{c c c}
				h=1 &: & 2 {\ \rm states}\\
				h=2 &: & 6 {\ \rm states}\\
				h=3 &: & 2 {\ \rm states}\\
		\end{array}
	\end{split}
\end{equation}
These twisted sector states  correspond to 10 additional gravitini (and 10 additional spin $1/2$ fermions), arising from the twisted sectors. To each one corresponds an additional supercharge, so that we have a total of 16 supercharges of positive helicity arising from the left-movers from all orbifold sectors.

From the worldsheet right-movers we have $8+8$ additional supercharges of opposite helicities
\begin{equation}
	\tilde Q_{+1/2} = e^{\tilde\phi/2+i\tilde H_0/2} \,\tilde S_8 \quad,\quad \tilde Q_{+1/2}' = e^{\tilde\phi/2-i\tilde H_0/2}\, \tilde C_8 \,,
\end{equation}
all of which are left intact by the asymmetric $\mathbb Z_4$ action. Therefore, the theory enjoys chiral $(N,M)=(24,8)$ supersymmetry and there are $128/8=16$ physical massless chiral multiplets of negative helicity.
We now see that the gravity multiplet actually contributes $24\cdot\frac{1}{2}-8\cdot\frac{1}{2}=8$ to the gravitational anomaly, while the 16 chiral multiplets of negative helicity yield  $-16\cdot\frac{8}{16}=-8$, as before. Therefore, the total anomaly coefficient indeed vanishes identically, as it should in type IIB theory, in accordance with modular invariance.

Let us further mention that the compactification of IIA string theory on the same asymmetric orbifold is essentially identical, and amounts to the exchange of the internal Weyl chiralities of the right-moving SO(8) spinors $\bar S_8 \leftrightarrow\bar C_8$ in both the $\mu=0$ and $\mu=1$ spectra. Indeed, a T-duality in one of the eight directions at this special point does not affect the lattice, but introduces an additional phase $(-1)^{\bar a\bar b}$ in the right-moving RNS partition function. Because the orbifold does not act on the right-movers, this simply flips the GSO projection of right-moving Ramond states $\bar a=1$ and exchanges their Weyl chiralities.

The type II theory on the asymmetric $\mathbb Z_4$ orbifold constructed here is the first `exotic' example we present in this work. Modulo chirality assignments,  it contains a total of 32 supercharges.  Theories with 32 supercharges are often thought to be ``trivial cases", in the sense that they only arise either directly in ten dimensions, or by compactification on simple torii. This is not the case with the (24,8) theory that we discussed here, which serves as an example of the rich structure of 2d constructions.



\subsection{Asymmetric $T^8/\mathbb Z_4$ in heterotic with $(16,0)$ supersymmetry}

The phenomenon of extra supercharges (or gravitini) arising from twisted sectors in fully asymmetric orbifold constructions in two spacetime dimensions is not particular to type II theories. As the discussion of previous section suggested, it is a property of the pure left-moving action of the twist. We therefore expect the same kind of supersymmetry ``enhancement" to occur also in heterotic theories compactified on the same asymmetric $T^8/\mathbb Z_4$ orbifold. 

To this end, we consider the heterotic ${\rm E}_8\times{\rm E}_8$ string theory compactified on this asymmetric orbifold, whose partition function reads
\begin{equation}
	\begin{split}
Z= \frac{1}{4}\sum_{h,g\in\mathbb Z_4} &\left[\frac{1}{2}\sum_{a,b}(-1)^{a+b+\mu ab}\, \frac{\vartheta[^{a+h/2}_{b+g/2}]^2 \vartheta[^{a-h/2}_{b-g/2}]^2}{\eta^4}\, e^{-i\pi hg/2}\right] \,\frac{\Gamma^{\rm asym}_{8,8}[^h_g]}{\eta^8\bar\eta^8}\,e^{-i\pi hg/2} \\
				\times  &\frac{1}{2}\sum_{k,\ell}\,\frac{\bar\vartheta[^k_\ell]^8}{\bar\eta^8}\times\frac{1}{2}\sum_{\rho,\sigma}\frac{\bar\vartheta[^\rho_\sigma]^8}{\bar\eta^8}  \,.
	\end{split}
\end{equation}
The theta function combinations in the second line simply reflect the free fermion representation of the anti-chiral ${\rm E}_8\times {\rm E}_8$ Kac-Moody lattice. As before, due to the asymmetry of the orbifold action, this construction is performed at the special fermionic point in moduli space. We will show that the theory has chiral $(16,0)$ supersymmetry with gauge group ${\rm SO}(16)\times {\rm E}_8\times {\rm E}_8$.

\subsubsection{Positive chirality spectrum ($\mu=0$ states)}
The analysis of the spectrum of states of positive spacetime chirality states, $\mu=0$, was outlined in the previous sections. As discussed there, the most convenient representation from which to analyse the massless (or even massive spectra) is the character expansion of the partition function in the generic orbifold sector $h=0,1,2,3$, while taking into account the splitting into spacetime bosonic-NS and spacetime fermionic-R states
\begin{equation}
	\begin{split}
	Z_{h}^{\rm NS} &= \left(\chi[^{0,h}_{-,1}]\,\hat\Gamma^{\rm asym}[^h_3]+\chi[^{0,h}_{-,3}]\,\hat\Gamma^{\rm asym}[^h_1]\right)\, (O_{16}+\bar S_{16})^2 \\
		Z_{h}^{\rm R} &= -\left(\chi[^{1,h}_{-,1}]\,\hat\Gamma^{\rm asym}[^h_3]+\chi[^{1,h}_{-,3}]\,\hat\Gamma^{\rm asym}[^h_1]\right) (\bar O_{16}+\bar S_{16})^2 
	\end{split}
\end{equation}
A close inspection of this decomposition shows the absence of massless physical (propagating) states with positive chirality in any sector $h=0,1,2,3$.

\subsubsection{Negative chirality spectrum ($\mu=1$ states)}
The same decomposition may be obtained for the negative chirality states, corresponding to the choice $\mu=1$,
\begin{equation}
	\begin{split}
	Z_{h}^{\rm NS} &= \left(\chi[^{0,h}_{-,1}]\,\hat\Gamma^{\rm asym}[^h_3]+\chi[^{0,h}_{-,3}]\,\hat\Gamma^{\rm asym}[^h_1]\right)\, (O_{16}+\bar S_{16})^2 \\
		Z_{h}^{\rm R} &= -\left(\chi[^{1,h}_{+,0}]\,\hat\Gamma^{\rm asym}[^h_0]+\chi[^{1,h}_{+,2}]\,\hat\Gamma^{\rm asym}[^h_2]\right) (\bar O_{16}+\bar S_{16})^2 
	\end{split}
\end{equation}
Like in the $\mu=0$ case, also in $\mu=1$ the bosons $a=0$ are again massive in all orbifold sectors $h=0,1,2,3$. Massless physical (propagating) states of negative spacetime chirality can, however, arise in the Ramond sector $a=1$. From the untwisted $h=0$ sector we find a total of 3696 fermionic states in the adjoint representation of the gauge group
\begin{equation}
	-\chi[^{1,0}_{+,0}]\,\hat\Gamma^{\rm asym}[^0_0] (\bar O_{16}+\bar S_{16})^2 \quad:\quad \textbf{6}_0 \times {\rm Adj}[ {\rm SO}(16)\times{\rm E}_8\times{\rm E}_8 ] \,.
\end{equation}
From the $h=1$ twisted sector we similarly find a total of 1744 fermionic states
\begin{equation}
	-\chi[^{1,1}_{+,2}]\,\hat\Gamma^{\rm asym}[^1_2]\,(\bar O_{16}+\bar S_{16})^2 \quad:\quad \tfrac{1}{4}[++++]\cdot\left[ (2q^{3/8})\cdot {\rm Adj}({\rm E}_8\times{\rm E}_8)+752q^{3/8}\bar q\right] \,.
\end{equation}
The $h=2$ twisted sector gives rise to a total of 4720 fermionic states
\begin{equation}
	-\chi[^{1,2}_{+,2}]\,\hat\Gamma^{\rm asym}[^2_2]\,(\bar O_{16}+\bar S_{16})^2 \quad:\quad 1\cdot\left[ (6q^{1/2})\cdot {\rm Adj}({\rm E}_8\times{\rm E}_8) + 1744 q^{1/2}\bar q\right]\,.
\end{equation}
Finally, from the $h=3$ twisted sector we find 1744 fermionic states
\begin{equation}
	-\chi[^{1,3}_{+,2}]\,\hat\Gamma^{\rm asym}[^3_2]\,(\bar O_{16}+\bar S_{16})^2 \quad:\quad \tfrac{1}{4}[----]\cdot\left[ (2q^{3/8})\cdot{\rm Adj}({\rm E}_8\times{\rm E}_8)+752q^{3/8}\bar q\right]\,.
\end{equation}
In total, there are 11904 massless propagating fermions of negative chirality, all of which are singlets with respect to the chiral $(16,0)$ supersymmetry.

\subsubsection{Supersymmetry charges}

To construct the vertex operators of supercharges, we again employ the covariant formalism. For heterotic theory,  the left-moving RNS sector requires the projection onto GSO-even $G=+1$ states, with GSO parity defined as
\begin{equation}
	G = e^{i\pi ( -Q_\phi + Q_0 + Q_1+ Q_2+Q_3+Q_4 )}\,.
\end{equation}
As in previous sections, we only display supercharges in the $+1/2$ super-ghost picture. From the untwisted sector, we have the familiar 6 invariant supercharges with positive spacetime helicity ($Q_0=+1/2$)
\begin{equation}
	Q_{+1/2} = e^{\phi/2+iH_0/2} \otimes \tfrac{1}{2}\left[ \begin{array}{c c c c}
												- & - & + & + \\
												- & + & - & + \\
												- & + & + & -\\
												+ & - & - & + \\
												+ & - & + & - \\
												+ & + & - & - \\
												\end{array}\right] \quad,\quad \textbf{6}_0\,.
\end{equation}
As in the asymmetric type II case, additional supercharges do arise from the twisted sectors
\begin{equation}
	\begin{split}
			e^{\phi/2+iH_0/2}\otimes\left(-\chi[^{1,h}_{+,2}]\,\hat\Gamma^{\rm asym}[^h_2]\right)\otimes \bar\partial X^\mu  \quad :\quad 2_{h=1}+6_{h=2}+2_{h=3}=10 \,.
	\end{split}
\end{equation}
In particular, we see that the twisted sectors give rise to 10 additional gravitini (and 10 additional spin $1/2$ fermions), so that we have a total of 16 supercharges of positive helicity arising from the left-movers from all orbifold sectors, and the theory enjoys chiral $(16,0)$ supersymmetry. 

\subsubsection{Anomaly and $B$-field tadpole}
Let us evaluate the gravitational anomaly coefficient for this construction. From the gravity multiplet, we have a contribution $16\cdot\frac{1}{2}=8$. On the other hand, from the 11904 negative chirality fermion supersymmetry singlets, we have a contribution $-11904\cdot\frac{1}{24}=-248$. The total anomaly coefficient is therefore found to be $-240$ and will be cancelled by the Green-Schwarz mechanism. As in the symmetric heterotic case, since the target space is two dimensional, this same coefficient will be also identified as the coefficient of the $B$-field tadpole that can be cancelled by introducing $n=240$ additional fundamental heterotic strings in the theory. 

This can be verified independently by performing a 1-loop computation, in order to fix the tadpole coefficient. From the partition function of the internal CFT, we can compute 
\begin{equation}
	A(\bar q,0,0)= 8\bar j(\bar \tau) = \frac{8}{\bar q}+5952 + \ldots \,,
\end{equation}
which, according to the notation of the symmetric heterotic orbifold of Section \ref{HetSymSection}, $\alpha=8$ and $\beta=5952$. The tadpole coefficient is then found to be $n=\beta/24-\alpha = 240$ and matches precisely the anomaly coefficient computed directly from the spectrum, as it should.



\subsection{Asymmetric $T^8/\mathbb Z_4$ in type IIB with shifts and $(20,8)$ supersymmetry}

In this section we consider type IIB string theory compactified on an asymmetric orbifold $T^8/\mathbb Z_4$ with a slightly modified action. Namely, in addition to the $\mathbb Z_4$ rotation of all 8 left-movers, the orbifold simultaneously acts as an order-2 shift on 4 right-moving $T^8$ coordinates. The partition function in this case reads
\begin{equation}
	\begin{split}
Z= \frac{1}{4}\sum_{h,g\in\mathbb Z_4} &\left[\frac{1}{2}\sum_{a,b}(-1)^{a+b+\mu ab}\, \frac{\vartheta[^{a+h/2}_{b+g/2}]^2 \vartheta[^{a-h/2}_{b-g/2}]^2}{\eta^4}\, e^{-i\pi hg/2}\right] \,\frac{{\tilde\Gamma}^{\rm asym}_{8,8}[^h_g]}{\eta^8\bar\eta^8}\,e^{-i\pi hg/2} \\
				\times  &\left[\frac{1}{2}\sum_{\bar a,\bar b}(-1)^{\bar a+\bar b+\mu \bar a\bar b}\, \frac{\bar\vartheta[^{\bar a}_{\bar b}]^4}{\bar\eta^4}\, \right]  \,,
	\end{split}
\end{equation}
where $\tilde\Gamma_{8,8}^{\rm asym}[^h_g]$ is the twisted/shifted $(8,8)$ lattice at the fermionic point, defined as follows
\begin{equation}
	\tilde\Gamma_{8,8}^{\rm asym}[^h_g] = \frac{1}{2}\sum_{\gamma,\delta}  \vartheta[^\gamma_\delta]^2\,\vartheta[^{\gamma+h/2}_{\delta+g/2}]^2\,\vartheta[^{\gamma-h/2}_{\delta-g/2}]^2\,\vartheta[^{\gamma-h}_{\delta-g}]\,\vartheta[^{\gamma+h}_{\delta+g}] \, \bar\vartheta[^\gamma_\delta]^4 \,\bar\vartheta[^{\gamma+h}_{\delta+g}]^4\,.
\end{equation}
Since the action of the orbifold involves an asymmetric rotation, the construction is again forced to lie at the fermionic point in the $T^8$ moduli space. We will see below that this theory has chiral $(20,8)$ supersymmetry. Similarly to the asymmetric type II construction of Section \ref{SecAsymIIB}, the trivial action of the orbifold on the right-moving RNS fermions implies that the IIA and IIB theories have essentially identical spectra and the same amount of supersymmetry.

We now turn to a brief discussion of the spectrum. All $\mu=0$ states are massive, in all orbifold sectors. The massless physical $\mu=1$ states (negative chirality) are identified as
\begin{equation}
	\begin{split}
	Z_{h=0}^{\rm R-R} = \chi[^{1,0}_{+,0}]\, \hat{\tilde\Gamma}[^0_0]\,\bar S_8 \quad &:\quad 6\times 8  \\
	Z_{h=0}^{\rm R-NS} =-\chi[^{1,0}_{+,0}]\, \hat{\tilde\Gamma}[^0_0]\,\bar V_8 \quad &:\quad 6\times 8\\
	Z_{h=2}^{\rm R-R} = \chi[^{1,2}_{+,2}]\,\hat{\tilde\Gamma}[^2_2]\,\bar S_8 \quad &:\quad 6\times 8\\
	Z_{h=2}^{\rm R-NS}=-\chi[^{1,2}_{+,2}]\, \hat{\tilde\Gamma}[^2_2]\,\bar V_8 \quad &:\quad 6\times 8\\
	\end{split}
\end{equation}
Notice that the sectors $h=1$ and $h=3$ are always massive. Therefore, we have 96 negative chirality R-R scalars and 96 negative chirality R-NS fermions.

\subsubsection{Supersymmetry charges}
As everywhere in this work, we will write (and count) all supercharges conventionally in the $+1/2$ superghost picture. Since the supercharges can only be counted in the full covariant framework, we will need to specify a conventional choice for the GSO projection. For type IIB, we will choose
\begin{equation}
	G= (-1)^{-Q_\phi+Q_0}\,G_{\rm int} \quad,\quad \tilde G = (-1)^{-\tilde Q_\phi+\tilde Q_0}\, \tilde G_{\rm int}\,,
\end{equation}
for the left and right moving GSO parity operators, and define also the``internal GSO parity" operators
\begin{equation}
	G_{\rm int} = (-1)^{Q_1+Q_2+Q_3+Q_4} \quad,\quad \tilde G_{\rm int} = (-1)^{\tilde Q_1+\tilde Q_2+\tilde Q_3+\tilde Q_4} \,.
\end{equation}

The general form of the positive and negative chirality supercharges (if they exist) is
\begin{equation}
	\begin{split}
	Q_{+1/2}^{(+)} &= e^{\phi/2+iH_0/2}\, \Sigma^r \ ,\quad {\rm positive\ chirality} \,,\\
	Q_{+1/2}^{(-)} &= e^{\phi/2-iH_0/2}\, \hat\Sigma^s \ ,\quad {\rm negative\ chirality} \,.
	\end{split}
\end{equation}
Here, $\Sigma^r$ is an operator in the internal CFT with conformal weight $(1/2,0)$, invariant under the orbifold action, and with internal GSO parity $G_{\rm int}=+1$.
Similarly, $\hat\Sigma^s$ is an operator in the internal CFT with conformal weight $(1/2,0)$, invariant under the orbifold action, but with the opposite internal GSO parity $G_{\rm int}=-1$. For the type IIB theory, identical expressions hold also for the right-moving supercharges with the analogous internal GSO parity requirements. 

Let us first examine the negative chirality supercharges arising from the left-movers. We are looking for orbifold invariant operators $\hat \Sigma$ with internal GSO $G_{\rm int}=-1$. For this theory, since the right-moving RNS fermion characters are completely factored from the remaining degrees of freedom, it suffices to look at the left-moving Ramond sector part of massless states with negative internal GSO projection, \emph{i.e.} the sector $a=1$, $\mu=0$. Since, as mentioned above, no massless states exist in the $\mu=0$ sector, we conclude that the left-movers do not give rise to any negative chirality supercharges.

Now we can turn to the positive chirality supercharges arising still from the left-movers. We are looking for orbifold invariant operators $\Sigma$ with positive internal GSO $G_{\rm int}=+1$. To find them, it is again sufficient \emph{in this particular theory} to consider the left-moving part in the Ramond sector of massless states with positive internal GSO projection, \emph{i.e.} the sector $a=1$, $\mu=1$. By looking at the $\mu=1$ massless spectrum, we can identify 2 such possibilities
\begin{equation}
	\begin{split}
		& h=0 \quad :\quad \chi[^{1,0}_{+,0}]\,\hat{\tilde \Gamma}[^0_0] \quad,\quad (6q^{1/2})(1+\ldots) \,,\\
		& h=2 \quad :\quad \chi[^{1,2}_{+,2}]\,\hat{\tilde \Gamma}[^2_2] \quad,\quad (1+\ldots)(6q^{1/2}) \,.\\
	\end{split}
\end{equation}
These 12  operators indeed have  conformal weight $(1/2,0)$, they are invariant under the orbifold action and are positive under the internal GSO, $G_{\rm int}=+1$. Therefore, the left-movers contribute a total of 12 positive chirality supercharges and no negative chirality supercharges, \emph{i.e.} $(12,0)$.

Now let us turn to supercharges arising from the right-movers. The supercharges will again have the generic form
\begin{equation}
	\begin{split}
	\tilde Q_{+1/2}^{(+)} &= e^{\tilde\phi/2+i\tilde H_0/2}\, \tilde\Sigma^r \ ,\quad {\rm positive\ chirality} \,,\\
	\tilde Q_{+1/2}^{(-)} &= e^{\tilde\phi/2-i\tilde H_0/2}\, \hat{\tilde\Sigma}^s \ ,\quad {\rm negative\ chirality} \,.
	\end{split}
\end{equation}
Let us first count the positive chirality supercharges. This amounts to finding invariant operators $\tilde\Sigma$, with conformal weight $(0,1/2)$, which have $\tilde G_{\rm int}=+1$. Since the orbifold does not act on the right-moving fermions, we can focus only on their contribution. The positive internal GSO requirement translates to $\mu=1$ and gives rise to the 8 states  $\tilde\Sigma \rightarrow \bar S_8$ and, hence, 8 supercharges of positive chirality.

For the negative chirality supercharges, we work in a similar way. We look for invariant operators $\hat{\tilde\Sigma}$ with conformal weight $(0,1/2)$ with $\tilde G_{\rm int}=-1$. The negative internal GSO requirement arises from the sector $\mu=0$ and again gives rise to 8 states $\hat{\tilde\Sigma}\rightarrow \bar C_8$, corresponding to 8 supercharges of negative chirality.

Therefore, the right-movers contribute a total of $(8,8)$ supercharges. Adding these together with the supercharges arising from the left-movers, we find that the full theory enjoys chiral $(20,8)$ supersymmetry.

As a check, we can verify the vanishing of the anomaly polynomial. There are 96 negative chirality R-R scalars, contributing $-96/24$ to the anomaly and an equal number of negative chirality R-NS fermions contributing $-96/48$. Since the theory has $(20,8)$ supersymmetry its gravity multiplet contributes $20/2-8/2$, and hence, we find that the anomaly indeed cancels.

The above discussion illustrates a peculiar phenomenon which, as we have seen, may arise in asymmetric orbifolds in two spacetime dimensions. In ordinary four-dimensional string constructions, supplementing an orbifold rotation with a translation does not affect the number of preserved supersymmetries of the theory. Typically, it is the rotation itself that determines which supercharges survive the projection, regardless of whether this rotation belongs to the space of ordinary coordinates such as in the case of partial breaking, or whether it only rotates the R-symmetry lattice, as in the case of full Scherk-Schwarz breaking of supersymmetry. In those cases, the effect of the translation is merely to render the action free and, for instance, to give mass to non-invariant gravitini instead of projecting them out of the theory. In the construction at hand, instead,  modifying the orbifold action as little as introducing an asymmetric shift does significantly affect the number of preserved supercharges -- specifically, those arising from the twisted sectors. 



\section{Two dimensional string theories with more than 32 supercharges}
\label{sec:more-than-32}

In closed string theory, one does not normally expect finite changes in the size and shape of the internal space to affect the number of preserved supersymmetries. Gravitini may become massless only at the boundary of moduli space. As we will see more clearly in the following, however, different choices for the shape of the $T^8$ lattice may result in different numbers of supersymmetries being preserved, after modding out by an asymmetric orbifold action. This  does not contradict our expectation that no supersymmetry enhancement may result from infinitesimal deformations in the bulk of moduli space, since the phenomenon we are describing is founded on the asymmetric nature of the orbifold. The would-be moduli responsible for continuously interpolating between two such theories preserving different numbers of supersymmetries are, in fact, absent.

In order to illustrate how different choices for the shape of the (8,8) lattice may lead to different numbers of supercharges, we shall examine two type II theories based on asymmetric $T^8/\mathbb Z_4\times\mathbb Z_4$ orbifolds. In both cases, the action of the orbifold on the worldsheet degrees of freedom will be the same from the point of view of the internal CFT of the auxiliary worldsheet fermions $y,w,\tilde y,\tilde w$ replacing the internal coordinates, but the way these elementary left- and right- moving degrees of freedom are then combined together in order to form ``coordinates" will be different. In other words, these two theories will differ in the choice of point in the perturbative moduli space of the Narain lattice, leading to different lattice symmetries.  Remarkably, we are going to show that a special factorised choice for the (8,8) lattice gives rise to a consistent 2d string theory with more than 32 supercharges!

\subsection{Asymmetric $T^8/\mathbb Z_4 \times \mathbb Z_4$ in type IIA/ IIB with $(16,16)$ or $(32,0)$ supersymmetry}

The first case we shall present involves type IIA/IIB string theory compactified on an asymmetric orbifold $T^8/\mathbb Z_4\times \mathbb Z_4$ which is, in a sense, a double copy of the asymmetric orbifold we have been considering so far. Namely, each of the $\mathbb Z_4$ orbifold factors independently act as asymmetric pure $\mathbb Z_4$ rotations \eqref{Z4orbaction} on the left (resp. right) movers. The IIB partition function explicitly reads
\begin{equation}
	\begin{split}
Z= \frac{1}{4}\sum_{h_1,g_1\in\mathbb Z_4}\frac{1}{4}\sum_{h_2,g_2\in\mathbb Z_4} &\left[\frac{1}{2}\sum_{a,b}(-1)^{a+b+\mu ab}\, \frac{\vartheta[^{a+h_1/2}_{b+g_1/2}]^2 \vartheta[^{a-h_1/2}_{b-g_1/2}]^2}{\eta^4}\, e^{-i\pi h_1g_1/2}\right] \,\frac{{\check\Gamma}^{\rm asym}_{8,8}[^{h_1, h_2}_{g_1,g_2}]}{\eta^8\bar\eta^8}\,e^{-i\pi (h_1g_1-h_2 g_2)/2} \\
				\times  &\left[\frac{1}{2}\sum_{\bar a,\bar b}(-1)^{\bar a+\bar b+\mu \bar a\bar b}\,  \frac{\bar\vartheta[^{\bar a+h_2/2}_{\bar b+g_2/2}]^2 \bar\vartheta[^{\bar a-h_2/2}_{\bar b-g_2/2}]^2}{\bar\eta^4}\, e^{+i\pi h_2g_2/2}\, \right]  \,,
	\end{split}
\end{equation}
where $\check\Gamma_{8,8}^{\rm asym}[^{h_1,h_2}_{g_1,g_2}]$ is the twisted $(8,8)$ lattice at the fermionic point,
\begin{equation}
	G_{IJ} = \frac{1}{2}\delta_{IJ} \quad,\quad B_{IJ} = \frac{1}{2}\epsilon_{IJ}\,,
	\label{GBpoint1}
\end{equation}
and is defined in terms of one-loop Jacobi theta constants as follows
\begin{equation}
	\check\Gamma_{8,8}^{\rm asym}[^{h_1,h_2}_{g_1,g_2}] = \frac{1}{2}\sum_{\gamma,\delta}  \vartheta[^\gamma_\delta]^2\,\vartheta[^{\gamma+h_1/2}_{\delta+g_1/2}]^2\,\vartheta[^{\gamma-h_1/2}_{\delta-g_1/2}]^2\,\vartheta[^{\gamma-h_1}_{\delta-g_1}]\,\vartheta[^{\gamma+h_1}_{\delta+g_1}] \, 
	\bar\vartheta[^\gamma_\delta]^2\,\bar\vartheta[^{\gamma+h_2/2}_{\delta+g_2/2}]^2\,\bar\vartheta[^{\gamma-h_2/2}_{\delta-g_2/2}]^2\,\bar\vartheta[^{\gamma-h_2}_{\delta-g_2}]\,\bar\vartheta[^{\gamma+h_2}_{\delta+g_2}] \,.
\end{equation}
Since the action of the orbifold involves asymmetric rotations, no continuous deformation away from the point \eqref{GBpoint1} is possible. The IIB theory has chiral $(32,0)$ supersymmetry, while for IIA one finds instead $(16,16)$. 

A detailed analysis of these theories is straightforward but technically heavy to display explicitly. We shall, therefore, discuss here only their salient features, starting with their massless physical spectra. In type IIA, all sectors are massive for both positive and negative chirality states. In type IIB, only RR sectors with negative chirality can give rise to massless states. The number of such massless states for each orbifold sector $(h_1,h_2)$ is found to be
\begin{equation}
\begin{split}
	(0,0) \ : \ 36 \quad,\quad (0,1) \ :\ 12 \quad,\quad (0,2) \ :\ 36 \quad,\quad (0,3)\ : \ 12 \,,\\
	(1,0) \ : \ 12 \quad,\quad (1,1) \ :\ 12 \quad,\quad (1,2) \ :\ 28 \quad,\quad (1,3)\ : \ 12 \,,\\
	(2,0) \ : \ 36 \quad,\quad (2,1) \ :\ 28 \quad,\quad (2,2) \ :\ 68 \quad,\quad (2,3)\ : \ 28 \,,\\
	(3,0) \ : \ 12 \quad,\quad (3,1) \ :\ 12 \quad,\quad (3,2) \ :\ 28 \quad,\quad (3,3)\ : \ 12 \,.\\
\end{split}
\end{equation}
In total, there are 384 massless RR states of negative spacetime chirality. The left movers give rise to a total of 16 supercharges, while the right movers contribute 16 more. As in the previous sections, this is due to the presence of extra chiral operators in the CFT that are used to construct gravitini, and which arise from twisted sectors according to the pattern
\begin{equation}
	\begin{split}
		\chi[^{1,0}_{+,0}] \, \hat\Gamma[^{0,0}_{0,0}] \quad &:\quad 6q^{1/2}(1+\ldots) \,,\\
		\chi[^{1,1}_{+,2}] \, \hat\Gamma[^{1,0}_{2,0}] \quad &:\quad (1q^{1/8})(2q^{3/8})\,,\\
		\chi[^{1,2}_{+,2}] \, \hat\Gamma[^{2,0}_{2,0}] \quad &:\quad (1+\ldots)(6q^{1/2})\,,\\
		\chi[^{1,3}_{+,2}] \, \hat\Gamma[^{3,0}_{2,0}] \quad &:\quad (1q^{1/8})(2q^{3/8}) \,.
	\end{split}
\end{equation}
As usual, the gravitini carry the same or opposite spacetime chirality depending on whether the theory is IIA or IIB and, therefore, we indeed count $(16,16)$ supercharges in IIA and chiral $(32,0)$ in IIB. No tadpole arises in the IIA case, while the IIB anomaly vanishes as expected
\begin{equation}
	\frac{N}{2} - \frac{384}{24} =0 \,,
\end{equation}
for $N=32$.


\subsection{Asymmetric $T^8/\mathbb Z_4 \times \mathbb Z_4$ in type IIA/ IIB with exotic $(24,24)$ or $(48,0)$}

We now come to the final example to be discussed in this work and also the most exotic one. To our knowledge, it is the first example of a consistent string construction with more than 32 supercharges. Following the discussion of the previous subsection, we consider again type II string theory compactified on the same asymmetric orbifold $T^8/\mathbb Z_4\times \mathbb Z_4$, with one of the $\mathbb Z_4$ orbifold factors acting asymmetrically as pure left-moving $\mathbb Z_4$ rotation \eqref{Z4orbaction}, while the other $\mathbb Z_4$ similarly rotates asymmetrically only the right movers. 

The novelty here is that the (8,8) lattice is now factorised into a purely holomorphic lattice of signature (8,0) times a purely anti-holomorphic (0,8) one. Before the orbifold action, the Narain lattice is expected to carry left- and right- moving modular weights $(4,4)$ and, hence, holomorphy and modularity uniquely associates this lattice with $({\rm E}_{8})_L \times ({\rm E}_{8})_R$, namely
\begin{equation}
	\Gamma_{8,8}(\tau,\bar\tau)=\Gamma_{E_8}(\tau) \times \bar\Gamma_{E_8}(\bar\tau) \,.
\end{equation}
This factorised enhancement point is achieved with a special choice for the $T^8$ metric and B-field, such as
\begin{equation}
	G_{IJ}= \left(
\begin{array}{rrrrrrrr}
 1 & -\frac{1}{2} & 0 & \frac{1}{2} & 0 & -\frac{1}{2} & 0 & 0 \\
 -\frac{1}{2} & 1 & -\frac{1}{2} & -\frac{1}{2} & \frac{1}{2} & 0 & 0 & \frac{1}{2} \\
 0 & -\frac{1}{2} & 1 & \frac{1}{2} & 0 & 0 & 0 & -\frac{1}{2} \\
 \frac{1}{2} & -\frac{1}{2} & \frac{1}{2} & 1 & 0 & -\frac{1}{2} & \frac{1}{2} & 0 \\
 0 & \frac{1}{2} & 0 & 0 & 1 & -\frac{1}{2} & 0 & \frac{1}{2} \\
 -\frac{1}{2} & 0 & 0 & -\frac{1}{2} & -\frac{1}{2} & 1 & -\frac{1}{2} & -\frac{1}{2} \\
 0 & 0 & 0 & \frac{1}{2} & 0 & -\frac{1}{2} & 1 & \frac{1}{2} \\
 0 & \frac{1}{2} & -\frac{1}{2} & 0 & \frac{1}{2} & -\frac{1}{2} & \frac{1}{2} & 1
\end{array}
\right)
\end{equation}
and
\begin{equation}
	B_{IJ} = \left(
\begin{array}{cccccccc}
 0 & -\frac{1}{2} & 0 & \frac{1}{2} & 0 & -\frac{1}{2} & 0 & 0 \\
 \frac{1}{2} & 0 & -\frac{1}{2} & -\frac{1}{2} & \frac{1}{2} & 0 & 0 & \frac{1}{2} \\
 0 & \frac{1}{2} & 0 & \frac{1}{2} & 0 & 0 & 0 & -\frac{1}{2} \\
 -\frac{1}{2} & \frac{1}{2} & -\frac{1}{2} & 0 & 0 & -\frac{1}{2} & \frac{1}{2} & 0 \\
 0 & -\frac{1}{2} & 0 & 0 & 0 & -\frac{1}{2} & 0 & \frac{1}{2} \\
 \frac{1}{2} & 0 & 0 & \frac{1}{2} & \frac{1}{2} & 0 & -\frac{1}{2} & -\frac{1}{2} \\
 0 & 0 & 0 & -\frac{1}{2} & 0 & \frac{1}{2} & 0 & \frac{1}{2} \\
 0 & -\frac{1}{2} & \frac{1}{2} & 0 & -\frac{1}{2} & \frac{1}{2} & -\frac{1}{2} & 0
\end{array}
\right)\,.
\end{equation}
It is straightforward to verify that with this choice, the matrix $G+B$ becomes upper triangular, and reflects the symmetry enhancement. Of course, transformations such as discrete reparametrisations may be employed to offer other equivalent bases for its construction. However, it is important to stress that the factorised point described here is physically distinct from the non-factorised one of eq.\eqref{GBpoint1} used in the previous section, even though both choices admit a CFT realisation in terms of auxiliary worldsheet fermions $y,w,\tilde y,\tilde w$. The difference between the two constructions from the point of view of these fermionic degrees of freedom lies in their different boundary condition assignments. By analysing this theory, we shall see that the enhancement ${\rm SO}(8)_{L,R} \to ({\rm E}_8)_{L,R}$ is responsible for increasing the number of extra gravitini and the full theory is shown to have more than 32 supercharges. 

The partition function of the IIB theory reads
\begin{equation}
	\begin{split}
Z= \frac{1}{4}\sum_{h_1,g_1\in\mathbb Z_4}\frac{1}{4}\sum_{h_2,g_2\in\mathbb Z_4} &\left[\frac{1}{2}\sum_{a,b}(-1)^{a+b+\mu ab}\, \frac{\vartheta[^{a+h_1/2}_{b+g_1/2}]^2 \vartheta[^{a-h_1/2}_{b-g_1/2}]^2}{\eta^4}\, e^{-i\pi h_1g_1/2}\right] \,\frac{\Gamma_{E_8}[^{h_1}_{g_1}]\,\bar\Gamma_{E_8}[^{h_2}_{g_2}]}{\eta^8\bar\eta^8}\,e^{-i\pi (h_1g_1-h_2 g_2)/2} \\
				\times  &\left[\frac{1}{2}\sum_{\bar a,\bar b}(-1)^{\bar a+\bar b+\mu \bar a\bar b}\,  \frac{\bar\vartheta[^{\bar a+h_2/2}_{\bar b+g_2/2}]^2 \bar\vartheta[^{\bar a-h_2/2}_{\bar b-g_2/2}]^2}{\bar\eta^4}\, e^{+i\pi h_2g_2/2}\, \right]  \,,				
	\end{split}
\label{PartExotic1}
\end{equation}
where $\Gamma_{E_8}[^{h}_{g}]$ is the purely holomorphic, twisted $E_8$ lattice, defined in terms of level one free fermion characters as follows
\begin{equation}
	\Gamma_{E_8}^{\rm asym}[^h_g] = \frac{1}{2}\sum_{\gamma,\delta}  \vartheta[^\gamma_\delta]^2\,\vartheta[^{\gamma+h/2}_{\delta+g/2}]^2\,\vartheta[^{\gamma-h/2}_{\delta-g/2}]^2\,\vartheta[^{\gamma-h}_{\delta-g}]\,\vartheta[^{\gamma+h}_{\delta+g}] \,.
\end{equation}
Similarly, the IIA partition function is simply obtained from \eqref{PartExotic1} by introducing an additional $(-1)^{\bar a\bar b}$ phase flipping the right-moving GSO projection of right-moving Ramond states.

In fact, the boundary condition assignments are such that the left- and right- moving CFTs completely factorise. Note that this is a global\footnote{A local factorisation would be true only up to boundary condition assignments.} factorisation. To illustrate this fact, we define
\begin{equation}
	f(\mu,\tau) \equiv \frac{1}{4}\sum_{h,g\in\mathbb Z_4}\frac{1}{2}\sum_{a,b}(-1)^{a+b+\mu ab+hg}\, \frac{\vartheta[^{a+h/2}_{b+g/2}]^2 \vartheta[^{a-h/2}_{b-g/2}]^2}{\eta^4}\,\frac{\Gamma_{E_8}[^{h}_{g}]}{\eta^8} \,,
\end{equation}
and observe that the partition function for type IIA and IIB can be expressed as the sesquilinear product
\begin{equation}
	Z_{\rm IIA} = f(\mu,\tau) \bar f(\mu+1,\bar\tau) \quad,\quad Z_{\rm IIB} = |f(\mu,\tau)|^2 \,.
\end{equation}
This holomorphic factorisation property has important consequences on the spectrum of the theory. It is straightforward to show that $f(\mu,\tau)$ is a holomorphic function on the compact Riemann surface $\mathcal F= {\rm SL}(2;\mathbb Z)\backslash \mathcal H^{*}$, with $\mathcal H^{*}$ being the Poincar\'e upper half-plane together with the point at infinity. Indeed, $f$ has no pole at $q=0$ and is, hence, a constant by the open mapping theorem
\begin{equation}
	f(\mu,\tau) = -12 (1-(-1)^\mu)\,.
	\label{MSDSident}
\end{equation}
In other words, the entire partition function factorises into a purely
holomorphic, times a purely anti-holomorphic part which, numerically,
turn out to be constants! The disappearance of the $\tau,\bar\tau$
dependence in the lightcone partition function is a consequence of
spacetime supersymmetry in the case $\mu=0$, or the consequence of an
MSDS spectral flow \cite{Florakis:2009sm,Florakis:2010ty} operating in
the massive spectrum of the theory in the case $\mu=1$.\footnote{While
  the MSDS spectral flow may be employed to explain \eqref{MSDSident}, we emphasize that the supercharges
  we discuss in this work are associated with genuine gravitini in
  two dimensions and are otherwise unrelated to the MSDS spectral flow that anyway operates on the non-supersymmetric negative chirality massive states ($\mu=1$).}

Using the identity \eqref{MSDSident}, it is easy to derive the spectrum of propagating massless states in both the type IIA and IIB theories under investigation. In type IIB, the $\mu=0$ sectors are massive, whereas the $\mu=1$ sectors give rise to $24^2=576$ massless RR scalars with negative spacetime chirality. In type IIA, all sectors are  massive.

We now turn our attention to the supercharges and work out the contribution of the left-movers only. The right-movers will give identical contributions, modulo their spacetime chirality that will either be the same as that of left-movers or opposite, depending on whether we are considering type IIB or type IIA, respectively. It is sufficient to look at the various orbifold sectors and identify all holomorphic operators of conformal weight $(\frac{1}{2},0)$ with Ramond boundary conditions on the left-movers 
\begin{equation}
	\begin{split}
			\chi[^{1,0}_{+,0}] \, \Gamma_{E_8}[^0_0] \quad &:\quad 6q^{1/2} (1+\ldots)  \,, \\
			\chi[^{1,1}_{+,2}] \, \Gamma_{E_8}[^1_2] \quad &:\quad (q^{1/8}) (4q^{3/8}) \,, \\
			\chi[^{1,2}_{+,2}]\, \Gamma_{E_8}[^2_2] \quad &:\quad (1)(10q^{1/2}) \,, \\
			\chi[^{1,3}_{+,2}]\, \Gamma_{E_8}[^3_2] \quad &:\quad (q^{1/8})(4q^{3/8}) \,.
	\end{split}
\end{equation}
Once these operators are tensored together with the superghost and longitudinal parts, they create the desired $(1,0)$ vertex operators of the supercharges. The left-movers contribute $(24,0)$ supersymmetries and the same is true for the right-movers, modulo chirality assignment. Therefore, we find that type IIB theory will enjoy chiral $(48,0)$ supersymmetry, whereas type IIA  has $(24,24)$.

It may appear surprising that we speak of supersymmetry in a theory where all propagating states are massive, such as in the case of type IIA, or for type IIB in the sector of positive chirality states. Nevertheless, the spectral flow in the massive $\mu=0$ sectors is the result of the supercharges identified above. Furthermore, the gravity multiplet does indeed organise itself according to the above counting of supersymmetries. As a check, consider the gravitational anomaly in the chiral type IIB theory
\begin{equation}
	\frac{N}{2} - \frac{576}{24} = 0 \,,
\end{equation}
which indeed vanishes for $N=48$.

Let us close this section with a comment about the vacuum energy. In the IIA case, the partition function numerically vanishes and, hence, so does the cosmological constant. The IIB case is more interesting and, in fact, may be calculated analytically
\begin{equation}
	 \Lambda_{\rm IIB} = -\int_{\mathcal F} \frac{d^2\tau}{\tau_2} \, (Z_{\mu=0}+Z_{\mu=1}) = -192\pi \,.
\end{equation}
In addition to non-trivial zero point energy, the theory also possesses non-trivial vacuum momentum and we refer to \cite{Dasgupta:1996yh} for further discussions of this point.


\section{Summary and further directions}

In this paper we have constructed several 2D left-right (a)symmetric orbifolds which lead to chiral theories 
with exotic $({\cal P},{\cal Q})$ supersymmetries. In some of the completely asymmetric cases an interesting and new
phenomenon occurs, namely the emergence of additional supercharges from the twisted sectors, such that the total number of supercharges is bigger than 32.
This happens if either in the left- or right-moving string sector the number of supercharges is bigger than 16, which so far never occurred in any 
$D\geq2$  string construction. The way how the two-dimensional orbifolds evade the common no-go theorem  that constrains the maximal number of supersymmetries  to
$\mathcal N\leq 8$ is the observation that in the sectors with ``too many" supersymmetries only massive propagating states exist.

A natural and very interesting question is whether the configurations
that we find have a higher dimensional description in some region of
parameter space. By virtue of being asymmetric orbifolds most of the
moduli of the internal space are projected out, so if possible at all,
we expect the appearance of a higher dimensional description to be
associated with regions of strong coupling in the IIA and heterotic
settings. Unfortunately finding a useful description of such strongly
coupled regions is far from straightforward: as discussed in
\cite{Dine:1997ji} the fact that we are dealing with an asymmetric
orbifold implies that the momentum modes in the eleventh-dimension are
accompanied by wrapped M5 branes, and the higher dimensional
formulation cannot simply be a weakly coupled supergravity. A related
fact is that the supersymmetry enhancements that we find are often
non-chiral, which precludes any construction from an ordinary circle
compactification of a higher dimensional theory.

Another question along similar lines is whether some, or all, of the
constructions we have obtained here can be understood in terms of
F-theory \cite{Vafa:1996xn}. While the formalism we use 
superficially looks very different than the tools one uses in
F-theory, there are known examples of F-theory embeddings of free
fermion models \cite{Berglund:1998eq,Berglund:1998rq}, so it is
certainly not impossible a priori that some dual exists. This is
particularly interesting in view of the recent activity in
constructing F-theory compactifications down to two dimensions
\cite{Schafer-Nameki:2016cfr,Apruzzi:2016iac,Lawrie:2016axq,Lawrie:2016rqe}. An
alternative approach for constructing a rich set of two-dimensional
theories are the ``brane-brick'' models recently studied in
\cite{Franco:2015tna,Franco:2015tya,Franco:2016nwv,Franco:2016qxh,Franco:2016fxm,Franco:2016tcm,Franco:2017cjj},
it would be equally interesting to see if any of the contructions here
have a geometric interpretation in this context.\footnote{Other
  interesting recent constructions of two dimensional theories can be
  found in
  \cite{Tatar:2015sga,Benini:2015bwz,Tatar:2017pnm,Gukov:2017zao}.}



\section*{Acknowledgements}

We wish to thank C.~Angelantonj, I.~Antoniadis, C.~Condeescu, S.~Ferrara, A-M.~Font, S.~Gukov and T.~Weigand for useful discussions. I.F.~would like  to further acknowledge the Theory Division at CERN for its warm hospitality during various stages of this work. The work of D.L.~is supported by the
ERC Advanced Grant No.~320045 ``Strings and Gravity".


\vskip2cm

{\bf \LARGE  \noindent{Appendices}}

  \appendix
\addtocontents{toc}{\protect\setcounter{tocdepth}{1}}
  
\section{Orbifold characters}\label{AppendixChar}

\subsection{Untwisted characters, bosons $h=0,a=0$}
We list below the untwisted bosonic characters defined in equation \eqref{DefChar}, including their $q$-expansion and the identification of their vertex operators.

\subsubsection{GSO odd}

For GSO odd characters $\xi=1$ and therefore, only odd values of $\lambda$ will be non-vanishing, i.e. $\lambda=1,3$. We therefore, have for $\lambda=1$
\begin{equation}
	q^{4/24}\,\chi[^{0,0}_{-,1}] = 4\sqrt{q}+32 q^{3/2} + 144 q^{5/2}+ \ldots
\end{equation}
The 4 states with conformal weight $1/2$ satisfying the conditions \eqref{GSOZ4cond} are identified as 
\begin{equation}
	 		\Psi^i = e^{i H_i} \quad, \quad \textbf{4}_{+1}\, {\rm of\, U(4)} \,,
\end{equation}
where we denote the U(1) charge of U(4) in the subscript, and we define it as $J=i\sum_j\partial H_j$.

The 32 states of weight $3/2$ can be identified as 
\begin{equation}
	\begin{split}
	\bar\Psi^i_{-1/2} \bar\Psi^j_{-1/2}\bar\Psi^k_{-1/2} \quad &: \ \bar{\textbf{4}}_{-3} \\
	 \Psi^i_{-1/2}\Psi^j_{-1/2}\bar\Psi^k_{-1/2} \quad &: \ \textbf{4}_{+1}+\textbf{20}_{+1}\\
	 \Psi^i_{-3/2} \quad &:\ \textbf{4}_{+1} 
	\end{split}
\end{equation}

For $\lambda=3$, we have
\begin{equation}
	q^{4/24}\,\chi[^{0,0}_{-,3}] = 4\sqrt{q}+32 q^{3/2} + 144 q^{5/2}+ \ldots
\end{equation}
However, now the 4 states with weight $1/2$ are identified as
\begin{equation}
	 		\bar\Psi^i = e^{-i H_i} \quad, \quad \bar{\textbf{4}}_{-1}\, {\rm of\, U(4)} \,.
\end{equation}
Similarly, for the 32 states of weight $3/2$, we find
\begin{equation}
	\begin{split}
	\Psi^i_{-1/2} \Psi^j_{-1/2} \Psi^k_{-1/2} \quad &: \ \textbf{4}_{+3} \\
	 \Psi^i_{-1/2}\bar\Psi^j_{-1/2}\bar\Psi^k_{-1/2} \quad &: \ \bar{\textbf{4}}_{-1}+\textbf{20}_{-1}\\
	\bar\Psi^i_{-3/2} \quad &:\ \bar{\textbf{4} }_{-1}
	\end{split}
\end{equation}

We thus see that the states in $\chi[^{0,0}_{-,3}]$ carry opposite charges with respect to those in $\chi[^{0,0}_{0,1}]$ and, therefore, correspond to the complex conjugate representations. 

\subsubsection{GSO even}
Now we deal with the case GSO even $\xi=0$, which means that only even values of $\lambda$ can arise. For the $\mathbb Z_4$ invariant states $\lambda=0$, we have
\begin{equation}
	q^{4/24}\,\chi[^{0,0}_{+,0}] = 1+16q+70q^2+ \ldots
\end{equation}
The 16 states with weight $1$ are identified as the currents generating the adjoint of U(4)
\begin{equation}
	\Psi^i_{-1/2} \bar\Psi^j_{-1/2} \quad : \ \textbf{15}_0+\textbf{1}  \,.
\end{equation}
For $\lambda=2$ one obtains the states transforming with a minus sign under $\mathbb Z_4$
\begin{equation}
	q^{4/24}\,\chi[^{0,0}_{+,2}] = 12q+64q^2+ \ldots
\end{equation}
These 12 states of weight $1$ can be recognized as 
\begin{equation}
	\begin{split}
	\Psi^i_{-1/2} \Psi^j_{-1/2} \quad &: \ \textbf{6}_{+2}  \,,\\
	 \bar\Psi^i_{-1/2}\bar\Psi^j_{-1/2} \quad &: \ \textbf{6}_{-2} \,.
	\end{split}
\end{equation}

\subsection{Untwisted characters, fermions $h=0,a=1$}
We move on to the untwisted fermionic characters. Again, we consider the GSO odd and even cases separately.

\subsubsection{GSO odd}

Since we are in the GSO odd case, $\xi=1$, the fermionic states transform necessarily with $\lambda=1$ or $\lambda =3$.
We have
\begin{equation}
	q^{4/24}\,\chi[^{1,0}_{-,1}] = 4\sqrt{q}+32 q^{3/2} + \ldots
\end{equation}
The 4 states with weight $1/2$ are identified as the states $e^{i Q \cdot H}$, with the weights given by
\begin{equation}
	\frac{1}{2}\left[ \begin{array}{cccc}
			- & + & + & + \\
			+& - & + & + \\
			+ & + & - & + \\
			+ & + & + & - \\
		\end{array}\right] \quad : \ \bar{\textbf{4}}_{+1}
\end{equation}
Similarly, we have 
\begin{equation}
	q^{4/24}\,\chi[^{1,0}_{-,3}] = 4\sqrt{q}+32 q^{3/2} + \ldots
\end{equation}
The 4 states with weight $1/2$ are now identified as the complex conjugate representation
\begin{equation}
	\frac{1}{2}\left[ \begin{array}{cccc}
			+ & - & - & - \\
			-& + & - & - \\
			- & - & + & - \\
			- & - & - & + \\
		\end{array}\right] \quad : \ \textbf{4}_{-1}
\end{equation}

\subsubsection{GSO even}

In the GSO even case, $\xi=0$, and the fermionic states transform necessarily with $\lambda=0$ or $\lambda =2$.
We have
\begin{equation}
	q^{4/24}\,\chi[^{1,0}_{+,0}] = 6\sqrt{q}+32 q^{3/2} + \ldots
\end{equation}
The 6 states with weight $1/2$ are identified as the states $e^{i Q \cdot H}$, with the weights given by
\begin{equation}
	\frac{1}{2}\left[ \begin{array}{cccc}
			+ & + & - & - \\
			+& - & + & - \\
			+ & - & - & + \\
			- & - & + & + \\
			- & + & - & + \\
			- & + & + & - \\
		\end{array}\right] \quad : \ \bar{\textbf{6}}_{0}
\end{equation}
Similarly, we have 
\begin{equation}
	q^{4/24}\,\chi[^{1,0}_{+,2}] = 2\sqrt{q}+32 q^{3/2} + \ldots
\end{equation}
The 2 states with weight $1/2$ are now identified with the two SU(4) singlets
\begin{equation}
	\frac{1}{2}\left[ \begin{array}{cccc}
			+ & + & + & + \\
			- & - & - & - \\
		\end{array}\right] \quad : \ \textbf{1}_{2}+\textbf{1}_{-2}
\end{equation}


\subsection{First twisted sector characters, $h=1$}

Working in a similar way, one may derive the character $q$-expansions and vertex operators also for the twisted sectors. For simplicity, we shall explicitly display here only the relevant $q$-expansions.
\begin{equation}
	\begin{split}
	& q^{4/24}\,\chi[^{0,1}_{-,1}] = 4q^{3/8}+28 q^{11/8} + \ldots \\
	& q^{4/24}\,\chi[^{0,1}_{-,3}] = 8q^{7/8}+56 q^{15/8} + \ldots \\
	& q^{4/24}\,\chi[^{0,1}_{+,0}] = q^{1/8}+17 q^{9/8} + \ldots \\
	& q^{4/24}\,\chi[^{0,1}_{+,2}] = 6q^{5/8}+38 q^{13/8} + \ldots \\
	& q^{4/24}\,\chi[^{1,1}_{-,1}] = 4q^{3/8}+28 q^{11/8} + \ldots \\
	& q^{4/24}\,\chi[^{1,1}_{-,3}] = 8q^{7/8}+56 q^{15/8} + \ldots \\	
	& q^{4/24}\,\chi[^{1,1}_{+,0}] = 6q^{5/8}+38 q^{13/8} + \ldots \\
	& q^{4/24}\,\chi[^{1,1}_{+,2}] = q^{1/8}+17 q^{9/8} + \ldots 
	\end{split}
\end{equation}


\subsection{Second twisted sector characters, $h=2$}

The character $q$-expansions for the $h=2$ twisted sector reads
\begin{equation}
	\begin{split}
	& q^{4/24}\,\chi[^{0,2}_{-,1}] = 4\sqrt{q}+32  q^{3/2} + \ldots \\
	& q^{4/24}\,\chi[^{0,2}_{-,3}] = 4 \sqrt{q}+32  q^{3/2} + \ldots \\
	& q^{4/24}\,\chi[^{0,2}_{+,0}] = 2 \sqrt{q}+32  q^{3/2} + \ldots \\
	& q^{4/24}\,\chi[^{0,2}_{+,2}] = 6\sqrt{q}+32 q^{3/2} + \ldots \\
	& q^{4/24}\,\chi[^{1,2}_{-,1}] = 4\sqrt{q}+32 q^{3/2} + \ldots \\
	& q^{4/24}\,\chi[^{1,2}_{-,3}] = 4\sqrt{q}+32 q^{3/2} + \ldots \\	
	& q^{4/24}\,\chi[^{1,2}_{+,0}] = 12 q +64 q^{2} + \ldots \\
	& q^{4/24}\,\chi[^{1,2}_{+,2}] = 1+16q + \ldots 
	\end{split}
\end{equation}


\subsection{Third twisted sector characters, $h=3$}

The character $q$-expansions for the $h=3$ twisted sector reads
\begin{equation}
	\begin{split}
	& q^{4/24}\,\chi[^{0,3}_{-,1}] = 8q^{7/8}+56 q^{15/8} + \ldots \\
	& q^{4/24}\,\chi[^{0,3}_{-,3}] = 4q^{3/8}+28 q^{11/8} + \ldots \\
	& q^{4/24}\,\chi[^{0,3}_{+,0}] = q^{1/8}+17 q^{9/8} + \ldots \\
	& q^{4/24}\,\chi[^{0,3}_{+,2}] = 6q^{5/8}+38 q^{13/8} + \ldots \\
	& q^{4/24}\,\chi[^{1,3}_{-,1}] = 8q^{7/8}+56 q^{15/8} + \ldots \\
	& q^{4/24}\,\chi[^{1,3}_{-,3}] = 4q^{3/8}+28 q^{11/8} + \ldots \\	
	& q^{4/24}\,\chi[^{1,3}_{+,0}] = 6q^{5/8}+38 q^{13/8} + \ldots \\
	& q^{4/24}\,\chi[^{1,3}_{+,2}] = q^{1/8}+17 q^{9/8} + \ldots 
	\end{split}
\end{equation}


\section{Lattice characters: Symmetric $\mathbb Z_4$ twist }\label{AppendixLat}

%
\subsection{Untwisted sector $h=0$}

We list the contributions arising from the untwisted sector
\begin{equation}
	\begin{split}
	&(q\bar q)^{8/24}\,\hat\Gamma^{\rm sym}[^0_{0}] = 1+32 q\bar q +\ldots \,,\\
	&(q\bar q)^{8/24}\,\hat\Gamma^{\rm sym}[^0_{1}] = 4 q+4\bar q  +\ldots\,,\\
	&(q\bar q)^{8/24}\,\hat\Gamma^{\rm sym}[^0_{2}] = 32 q\bar q  +\ldots\,,\\
	&(q\bar q)^{8/24}\,\hat\Gamma^{\rm sym}[^0_{3}] = 4 q+4\bar q  +\ldots\,.\\
	\end{split}
\end{equation}
It is straightforward to identify these lattice states at the generic point in moduli space. Strictly speaking, the lowest ones depicted here are not lattice states, but come from the Cartan directions. If we complexify the 8 coordinates into $Z^i$ and $\bar Z^i$, with $i=1,2,3,4$ then there are $4^2=16$ oscillators $\partial Z^i \bar\partial \bar Z^j$ and similarly another 16 oscillators given by $\partial \bar Z^i \bar\partial Z^j$ which are invariant. Similarly, the 4 non-level matched oscillators $\partial Z^i$ and similarly the 4 right-moving ones $\bar\partial Z^j$ correspond to the contributions of $\hat\Gamma[^0_1]$. The remaining contributions are analyzed in a similar fashion.

\subsection{Twisted sector $h=1$}
In the $h=1$ sector, the  contributions are
\begin{equation}
	\begin{split}
	&(q\bar q)^{8/24}\,\hat\Gamma^{\rm sym}[^1_{0}] = 16 q^{3/8} \bar q^{3/8} +256  q^{5/8} \bar q^{5/8}+1600  q^{7/8} \bar q^{7/8}+\ldots  \,,\\
	&(q\bar q)^{8/24}\,\hat\Gamma^{\rm sym}[^1_{1}] = 64  q^{3/8}\bar q^{5/8}+640 q^{5/8} \bar q^{7/8} +3840 q^{7/8} \bar q^{9/8} +\ldots \,,\\
	&(q\bar q)^{8/24}\,\hat\Gamma^{\rm sym}[^1_{2}] = 160  q^{3/8}\bar q^{7/8}+1536  q^{5/8}\bar q^{9/8}+160  q^{7/8}\bar q^{3/8}+\ldots \,,\\
	&(q\bar q)^{8/24}\,\hat\Gamma^{\rm sym}[^1_{3}] = 384 q^{3/8}\bar q^{9/8} +64  q^{5/8}\bar q^{3/8}+640  q^{7/8}\bar q^{5/8}+\ldots  \,.\\
	\end{split}
\end{equation}

\subsection{Twisted sector $h=2$}
In the $h=2$ sector, the  contributions are
\begin{equation}
	\begin{split}
	&(q\bar q)^{8/24}\,\hat\Gamma^{\rm sym}[^2_{0}] = 136  \sqrt{q}\sqrt{\bar q}+8192  q\bar q +\ldots \,,\\
	&(q\bar q)^{8/24}\,\hat\Gamma^{\rm sym}[^2_{1}] = 1024  \sqrt{q}\bar q+1024 q\sqrt{\bar q}  +\ldots \,,\\
	&(q\bar q)^{8/24}\,\hat\Gamma^{\rm sym}[^2_{2}] = 120  \sqrt{q}\sqrt{\bar q}+8192  q\bar q +\ldots\,,\\
	&(q\bar q)^{8/24}\,\hat\Gamma^{\rm sym}[^2_{3}] =  1024  \sqrt{q}\bar q+1024 q\sqrt{\bar q}+\ldots \,.\\
	\end{split}
\end{equation}

\subsection{Twisted sector $h=3$}
In the $h=3$ sector, the  contributions are
\begin{equation}
	\begin{split}
	&(q\bar q)^{8/24}\,\hat\Gamma^{\rm sym}[^3_{0}] = 16 q^{3/8} \bar q^{3/8} +256  q^{5/8} \bar q^{5/8}+1600  q^{7/8} \bar q^{7/8} +\ldots \,,\\
	&(q\bar q)^{8/24}\,\hat\Gamma^{\rm sym}[^3_{1}] = 64  q^{5/8}\bar q^{3/8}+640 q^{7/8} \bar q^{5/8} +3840 q^{9/8} \bar q^{7/8}+\ldots  \,,\\
	&(q\bar q)^{8/24}\,\hat\Gamma^{\rm sym}[^3_{2}] = 160  q^{3/8}\bar q^{7/8}+1536  q^{5/8}\bar q^{9/8}+160  q^{7/8}\bar q^{3/8}+1536  q^{9/8}\bar q^{5/8}+\ldots \,,\\
	&(q\bar q)^{8/24}\,\hat\Gamma^{\rm sym}[^3_{3}] = 384 q^{3/8}\bar q^{5/8} +64  q^{5/8}\bar q^{7/8}+640  q^{7/8}\bar q^{9/8}  +\ldots\,.\\
	\end{split}
\end{equation}

\section{Lattice characters: Asymmetric $\mathbb Z_4$ twist }\label{AppendixLatAsym}

%
%
\subsection{Untwisted sector $h=0$}

We list the contributions arising from the untwisted sector
\begin{equation}
	\begin{split}
	&(q\bar q)^{8/24}\,\hat\Gamma^{\rm asym}[^0_{0}] =1+28 q+120 \bar q+64 \sqrt{q} \sqrt{\bar q}+11552  q\bar q  \,,\\
	&(q\bar q)^{8/24}\,\hat\Gamma^{\rm asym}[^0_{1}] = 64 \sqrt{q} \sqrt{\bar q}+32 q+12032  q\bar q \,,\\
	&(q\bar q)^{8/24}\,\hat\Gamma^{\rm asym}[^0_{2}] = 64 \sqrt{q} \sqrt{\bar q}+28 q+11552  q\bar q \,,\\
	&(q\bar q)^{8/24}\,\hat\Gamma^{\rm asym}[^0_{3}] =  64 \sqrt{q} \sqrt{\bar q}+32 q+12032  q\bar q\,.\\
	\end{split}
\end{equation}

\subsection{Twisted sector $h=1$}

In the  first twisted sector $h=1$, we have
\begin{equation}
	\begin{split}
	&(q\bar q)^{8/24}\,\hat\Gamma^{\rm asym}[^1_{0}] = 32  q^{3/8}\bar q^{1/2}+20 q^{7/8}+7520  q^{7/8}\bar q\,,\\
	&(q\bar q)^{8/24}\,\hat\Gamma^{\rm asym}[^1_{1}] = 8 q^{5/8}+3008  q^{5/8}\bar q+768  q^{9/8}\sqrt{\bar q}\,,\\
	&(q\bar q)^{8/24}\,\hat\Gamma^{\rm asym}[^1_{2}] = 2 q^{3/8}+752  q^{3/8}\bar q+320  q^{7/8}\sqrt{\bar q} \,,\\
	&(q\bar q)^{8/24}\,\hat\Gamma^{\rm asym}[^1_{3}] = 128  q^{5/8}\sqrt{\bar q}+48 q^{9/8}+18048  q^{9/8}\bar q\,.\\
	\end{split}
\end{equation}

\subsection{Twisted sector $h=2$}

In the second twisted sector $h=2$,
\begin{equation}
	\begin{split}
	&(q\bar q)^{8/24}\,\hat\Gamma^{\rm asym}[^2_{0}] =2 \sqrt{q}+1264  \sqrt{q}\bar q+512  q\sqrt{\bar q}\,,\\
	&(q\bar q)^{8/24}\,\hat\Gamma^{\rm asym}[^2_{1}] =64  \sqrt{q}\sqrt{\bar q}+32 q+12032  q\bar q\,,\\
	&(q\bar q)^{8/24}\,\hat\Gamma^{\rm asym}[^2_{2}] =6 \sqrt{q}+1744  \sqrt{q}\bar q+512  q\sqrt{\bar q}\,,\\
	&(q\bar q)^{8/24}\,\hat\Gamma^{\rm asym}[^2_{3}] =64 \sqrt{q} \sqrt{\bar q}+32 q+12032  q\bar q\,.\\
	\end{split}
\end{equation}

\subsection{Twisted sector $h=3$}

Finally, the contributions of the third twisted sector $h=3$ read
\begin{equation}
	\begin{split}
	&(q\bar q)^{8/24}\,\hat\Gamma^{\rm asym}[^3_{0}] =32  q^{3/8}\sqrt{\bar q}+20 q^{7/8}+7520  q^{7/8}\bar q\,,\\
	&(q\bar q)^{8/24}\,\hat\Gamma^{\rm asym}[^3_{1}] =128  q^{5/8}\sqrt{\bar q}+48 q^{9/8}+18048  q^{9/8}\bar q\,,\\
	&(q\bar q)^{8/24}\,\hat\Gamma^{\rm asym}[^3_{2}] =2 q^{3/8}+752  q^{3/8}\bar q+320  q^{7/8}\sqrt{\bar q}\,,\\
	&(q\bar q)^{8/24}\,\hat\Gamma^{\rm asym}[^3_{3}] =8 q^{5/8}+3008  q^{5/8}\bar q+768  q^{9/8}\sqrt{\bar q}\,.\\
	\end{split}
\end{equation}


\bibliographystyle{JHEP}
\bibliography{refs}


\end{document}